\documentstyle[12pt]{article}
\def\ln{{\rm ln}}

\def\a{\begin{eqnarray}}
\def\b{\end{eqnarray}}
\def\0{\nonumber}
\def\ba{\begin{array}}
\def\ea{\end{array}}
\def\noal{\noalign{\vskip10pt}}

\def\q{{\bar{\cal Q}}}

\def\al{{\alpha}}
\def\lm{{\lambda}}

\def\cm{{\cal M}}

\def\tr{{\rm Tr}}
\renewcommand{\theequation}{\thesection.\arabic{equation}}

\newlength{\extraspace}
\newlength{\extraspaces}
\newcounter{dummy}
\newcommand{\ai}{
\addtocounter{equation}{1}
\setcounter{dummy}{\value{equation}}
\setcounter{equation}{0}
\renewcommand{\theequation}{\thesection.\arabic{dummy}\alph{equation}}
\begin{eqnarray}
\addtolength{\abovedisplayskip}{\extraspaces}
\addtolength{\belowdisplayskip}{\extraspaces}
\addtolength{\abovedisplayshortskip}{\extraspace}
\addtolength{\belowdisplayshortskip}{\extraspace}}
\newcommand{\bj}{
\end{eqnarray}
\setcounter{equation}{\value{dummy}}
\renewcommand{\theequation}{\thesection.\arabic{equation}}}
\def\d{{\partial}}

\newcommand{\ddlm}[1]{{\partial \over \partial \lm_{#1}}}

\def\ddt2{{{\d}\over{\d t_2}}}

\newcommand{\bac}{\begin{array}{c}}
\newcommand{\bacc}{\begin{array}{cc}}
\newcommand{\baccc}{\begin{array}{ccc}}
\newcommand{\barcl}{\begin{array}{rcl}}
\newcommand{\bacccc}{\begin{array}{cccc}}
\newcommand{\baccccc}{\begin{array}{ccccc}}
\newcommand{\baccccccc}{\begin{array}{ccccccc}}
\newcommand{\barclcrcl}{\begin{array}{rclcrcl}}
\newcommand{\bacl}{\begin{array}{cl}}
\newcommand{\bal}{\begin{array}{l}}
\newcommand{\bacll}{\begin{array}{cll}}
\begin{document}
\begin{flushright}
SISSA-ISAS 189/94/EP\\
UTHEP--695\\
hep-th/9412006\\
(new version)
\end{flushright}
\vskip0.5cm
\centerline{\LARGE\bf Exact correlators of two--matrix models}
\vskip0.3cm
\centerline{\large  L.Bonora, C.P.Constantinidis}
\centerline{International School for Advanced Studies (SISSA/ISAS)}
\centerline{Via Beirut 2, 34014 Trieste, Italy}
\centerline{INFN, Sezione di Trieste.  }
\vskip0.5cm
\centerline{\large C.S.Xiong}
\centerline{Department of Physics, University of Tokyo}
\centerline{Bunkyo--ku, Tokyo 113, Japan.}
\vskip5cm
\abstract{We compute exact solutions of two--matrix models, i.e.
detailed genus by genus expressions for the correlation functions of these
theories, calculated without any approximation. We distinguish between
two types of models, the unconstrained and the constrained ones.
Unconstrained two--matrix models represent perturbations of $c=1$ string
theory, while the constrained ones correspond to topological field theories
coupled to topological gravity. Among the latter we treat in particular
detail the ones based on the KdV and on the Boussinesq hierarchies.}

\vfill\eject

\section{Introduction}

\setcounter{equation}{0}
\setcounter{footnote}{0}

Matrix models represent sums over discretizations of Riemann surfaces,
possibly with some additional interactions. They are believed to provide a
(discrete) description of two dimensional gravity coupled to
matter. One--matrix models have been widely investigated,
but  their content is rather poor. The structure
of multi--matrix models is much richer but not yet known
as carefully as for one--matrix models (see \cite{Douglas},
\cite{is},\cite{ising},
\cite{G},\cite{GN},\cite{tada},\cite{DEB},
\cite{R},\cite{AS},\cite{MM},\cite{Kostov},\cite{PR},\cite{DKK},\cite{Stau}).

In this paper we concentrate on two--matrix models with bilinear coupling
and show how to find {\it exact solutions}.
By exact solutions we mean detailed genus by genus expressions
for the correlation functions of these
theories, calculated without any approximation. In particular, we do not limit
ourselves to exhibiting recursion relations which allow one to
compute correlators, but develop techniques to explicitly solve them.

The idea at the basis of our treatment of two--matrix models, outlined in
\cite{BX1}, is to transform
the initial functional integral problem into a discrete integrable linear
system
subject to some constraints (the coupling conditions). We end up in this way
with the discrete Toda lattice hierarchy. The latter
underlies all our calculations: our aim is to compute
the correlation functions (CF's) of each model, which in turn may be expressed
in terms of the integrable flows of the Toda hierarchy, subject to the coupling
conditions. This is the general setting for {\it unconstrained}
two--matrix models (which simply denote two--matrix models defined by specific
potentials without any further conditions).

We will also consider other models, obtained by suitably
constraining the previous (unconstrained) models. Their correlators
can be expressed either in terms of the flows of a reduced differential
hierarchy or in terms of suitably renormalized flows of the discrete
Toda hierarchy. In the process of solving these models we find a new
way of extracting integrable differential hierarchies from the Toda lattice
hierarchies.

Unconstrained two--matrix models describe various perturbations of
$c=1$ string theory at the self--dual point. Constrained models
correspond to well-known topological field theories coupled to topological
gravity.

The paper is organized as follows. In section 2 we collect the results
obtained in previous papers which will be necessary in the following.
In section 3 we discuss and calculate CF's of {\it unconstrained models}.
Section 4 is a short summary of how to obtain differential hierarchies
from the Toda lattice flows and reduced hierarchies via hamiltonian reduction.
In section 5, from the reduced hierarchies we construct a series of models,
named reduced models, which have a topological field theory meaning:
we show in particular how to compute all genus correlators. The reduced models
turn out to be imbedded into the constrained two--matrix models,
which are studied and solved in section 6.

\section{General properties of two--matrix models.}

\setcounter{equation}{0}
\setcounter{footnote}{0}

The model of two Hermitean $N \times N$ matrices $M_1$ and $M_2$,
is introduced in terms of the partition function
\a
Z_N(t,c)=\int dM_1dM_2 e^{trU},\quad\quad
U=V_1 + V_2 + g M_1 M_2\label{Zo}
\b
with potentials
\a
V_{\al}=\sum_{r=1}^{p_\al}\bar t_{\al,r}M_{\al}^r\,\qquad \al=1,2.\label{V}
\b
where $p_\al$ are finite numbers. These potentials define the model.
We denote by
${\cal M}_{p_1,p_2}$ the corresponding two--matrix model.

We are interested in computing correlation functions (CF's)
of the operators
\a
\tau_k=tr M_1^k,\qquad \sigma_k=tr M_2^k ,\qquad
\forall k,\qquad \chi= tr(M_1M_2)\0
\b
For this reason we complete the above model by replacing
(\ref{V}) with the more general potentials
\a
V_\al = \sum_{r=1}^\infty t_{\al,r} M_\al^r, \qquad \al =1,2\label{Vgen}
\b
where $t_{\al , r} \equiv \bar t_{\al,r}$ for $r\leq p_\al$. The CF's are
defined by
\a
< \tau_{r_1}\ldots \tau_{r_n}\sigma_{s_1}\ldots \sigma_{s_m}\chi^l> =
\frac {\d^{n+m+l} }{\d t_{1,r_1}\ldots \d t_{1,r_n}\d t_{2,s_1}\ldots
\d t_{2,s_m}\d g^l} \ln Z_N(t,g)\label{CF}
\b
where, in the RHS, all the $t_{\al,r}$ except $\bar t_{\al,r}$ are set to zero.

In other words we have embedded the original couplings $\bar t_{\al,r}$ into
two infinite sets of couplings. Therefore we have two types of couplings.
The first type consists of those couplings (the barred ones)
that define the model: they
represent the true {\it dynamical} parameters of the theory; they
are kept non-vanishing throughout the calculations.
The second type encompasses the remaining couplings, which
are introduced only for computational purposes and are set
to zero in formulas like (\ref{CF}).
In terms of ordinary field theory the former are analogous to the
interaction couplings, while the
latter correspond to external sources (coupled to composite operators).
{}From now on we will not
make any formal distinction between them.
Case by case we will specify which are the interaction couplings and which
are the external ones. Finally, it is sometime convenient to consider
$N$ on the same footing as the couplings and to set $t_{1,0}\equiv t_{2,0}
\equiv N$.

The path integral
(\ref{Zo}) is an ordinary integral in the matrix entries and it is
certainly well defined as long as a negative coupling $\bar t_{\al , r}$
with highest even $r$ guarantees that the measure is square--integrable and
decreases more than polynomially at infinity.
For the time being let us suppose that this is so.
Later on we will extend our problem to a larger coupling space.

We briefly recall the ordinary procedure to calculate the partition
function. It consists of three steps \cite{BIZ},\cite{IZ2},\cite{M}:
$(i)$ one integrates out the angular part so that only the
integrations over the eigenvalues are left;
$(ii)$ one introduces the orthogonal monic polynomials
\a
\xi_n(\lambda_1)=\lambda_1^n+\hbox{lower powers},\qquad\qquad
\eta_n(\lambda_2)=\lambda_2^n+\hbox{lower powers}\0
\b
which satisfy the orthogonality relations
\a
\int  d\lambda_1d\lambda_2\xi_n(\lambda_1)
e^{\mu(\lambda_1,\lambda_2)}
\eta_m(\lambda_2)=h_n(t,c)\delta_{nm}\label{orth1}
\b
where
\a
\mu(\lambda_1,\lambda_2) = {V_1(\lm_1)+V_2(\lm_2)+c\lm_1\lm_2}\0
\b

$(iii)$, using the orthogonality relation (\ref{orth1}) and the properties
of the Vandermonde determinants, one can easily
calculate the partition function
\a
Z_N(t,c)={\rm const}~N!\prod_{i=0}^{N-1}h_i\label{parti1}
\b

\subsection{From path integral to integrable systems}.

{}From (\ref{parti1}) we see that knowing the partition function means knowing
the coefficients $h_n(t,c)$.

The crucial point, from our point of view, is that the information
concerning the latter can be encoded in 1) a suitable linear system
subject to certain 2) coupling conditions,
together with 3) relations that allows us to reconstruct $Z_N$.
But before we pass to these three elements we need some convenient notations.
For any matrix $M$, we define
\a
\cm=H^{-1} MH,\qquad H_{ij}=h_i\delta_{ij},\qquad \bar M_{ij} = M_{ji},
\qquad M_l(j)\equiv M_{j,j-l}.\0
\b
As usual we introduce the natural gradation
\a
deg[E_{ij}] = j -i, \qquad {\rm where} \qquad (E_{i,j})_{k,l}= \delta_{i,k}
\delta_{j,l}\0
\b
and, for any given matrix $M$, if all its non--zero elements
have degrees in the interval $[a,b]$, then we will simply
write: $M\in [a,b]$. Moreover $M_+$ will denote the upper triangular
part of $M$ (including the main diagonal), while $M_-=M-M_+$. We will write
\a
{\rm Tr} (M)= \sum_{i=0}^{N-1} M_{ii}\0
\b
The latter operation will be referred to as taking the {\rm finite trace}.

Next we pass from the basis of orthogonal polynomials
to the basis of orthogonal functions
\a
\Psi_n(\lambda_1)=e^{V_1(\lambda_1)}\xi_n(\lambda_1),
\qquad
\Phi_n(\lambda_2)=e^{V_2(\lambda_2)}\eta_n(\lambda_2).\0
\b
The orthogonality relation (\ref{orth1}) becomes
\a
\int d\lm_1 d\lm_2\Psi_n(\lambda_1)e^{c\lm_1\lm_2}
\Phi_m(\lambda_2)=\delta_{nm}h_n(t,c).\label{orth2}
\b
We will denote by $\Psi$ the semi--infinite column vector
$(\Psi_0,\Psi_1,\Psi_2,\ldots,)^t$ and by $\Phi$ the vector
$(\Phi_0,\Phi_1,\Phi_2,\ldots,)^t$.

Then we introduce the following $Q$--type matrices
\a
\int d\lm_1 d\lm_2\Psi_n(\lambda_1)
\lm_{\al}e^{c\lm_1\lm_2}
\Phi_m(\lambda_2)\equiv Q_{nm}(\al)h_m=\q_{mn}(\al)h_n,\quad
\al=1,2.\label{Qalpha}
\b
Both $Q(1)$ and $\q(2)$ are Jacobi matrices: their pure upper triangular
part is $I_+=\sum_i E_{i,i+1}$.

Beside the above $Q$ matrices, we will need two $P$--type matrices, defined by
\a
&&\int d\lm_1 d\lm_2\Bigl(\ddlm 1 \Psi_n(\lambda_1)\Bigl)
e^{c\lm_1\lm_2}\Phi_m(\lambda_2)\equiv P_{nm}(1)h_m\label{P(1)}\\
&&\int  d\lambda_1d\lambda_2\Psi_n(\lambda_1)e^{c\lm_1\lm_2}
\Bigl(\ddlm 2 \Phi_m(\lambda_2)\Bigl)\equiv P_{mn}(2)h_n\label{P(2)}
\b
For later use we also introduce
\a
&&\int d\lm_1 d\lm_2\Bigl(\ddlm 1 \xi_n(\lambda_1)\Bigl)
e^{V_1(\lm_1)+V_2(\lm_2)+c\lm_1\lm_2}\eta_m(\lambda_2)\equiv
P^\circ_{nm}(1)h_m\label{P(1)'}\\
&&\int  d\lambda_1d\lambda_2\xi_n(\lambda_1)e^{V_1(\lm_1)+V_2(\lm_2)
+c\lm_1\lm_2}
\Bigl(\ddlm 2 \eta_m(\lambda_2)\Bigl)\equiv P^\circ_{mn}(2)h_n\label{P(2)'}
\b

Let us come now to the three elements announced above.

1) ${\underline {\it Coupling~~ conditions}}$.
\noindent The two matrices (\ref{Qalpha}) we
introduced above are not completely independent. More precisely both
$Q(\alpha)$'s can be expressed in terms of only one of them and
one matrix $P$.
Expressing the trivial fact that the integral of the total derivative of the
integrand in eq.(\ref{orth2}) with respect to $\lm_1$ and $\lm_2$
vanishes, we can easily derive the constraints or {\it coupling conditions}
\a
P(1)+c Q(2)=0,\qquad\quad
c Q(1)+\bar{\cal P}(2)=0,\label{coupling}
\b
{}From the coupling conditions it follows at once that, if we set to zero the
external couplings,
\a
Q(\al)\in[-m_{\al}, n_{\al}],\qquad \al=1,2\0
\b
where
\a
m_1=p_2-1, \qquad\quad m_2=1 \0
\b
and
\a
n_1=1, \qquad\quad n_2=p_1-1\0
\b
where $p_\al$, $\al =1,2$ is the highest order of the interacting part of
the potential $V_\al$ (see above).

2) ${\underline {\it The~ associated~~ linear~~ systems}}$.
\noindent The derivation of the  linear systems associated to our matrix model
is very simple.  We take the derivatives of eqs.(\ref{orth2})
with respect to the time parameters $t_{\al,r}$, and use
eqs.(\ref{Qalpha}).  We get in this way the time evolution of $\Psi$, or
{\it discrete linear system I}:
\a
\left\{\ba{ll}
Q(1)\Psi(\lambda_1)=\lambda_1\Psi(\lambda_1),& \\\noal
{\partial\over{\partial t_{1,k}}}\Psi(\lambda_1)=Q^k(1)_+
\Psi(\lambda_1),\\\noal
{\partial\over{\partial t_{2,k}}}\Psi(\lambda_1)=-Q^k(2)_-
\Psi(\lambda_1),\\\noal
{\partial\over{\partial\lm}}\Psi(\lambda_1)=P(1)\Psi(\lm_1).&
\ea\right.\label{DLS1}
\b
The corresponding consistency conditions are
\ai
&&[Q(1), ~~P(1)]=1\label{CC11}\\
&&{\partial\over{\partial t_{\al,k}}}Q(1)=[Q(1),~~Q^k(\al)_-],\qquad \alpha
=1,2\label{CC12}
\bj
In a similar way we can get the time evolution of $\Phi$ via a
{\it discrete linear system II}, whose consistency conditions are
\ai
&&[\q(2),~~P(2)]=1,\label{CC21}\\
&&{\partial\over{\partial t_{\al,k}}}Q(2)=[Q^k(\al)_+,~~Q(2)]\label{CC22}
\bj
We recall that one can write
down flows for $P(1)$ and $P(2)$, but they will not be used in this paper.

3) ${\underline {Reconstruction formulae}}$.
\noindent
The third element announced above is the link between the
quantities that appear in the linear system and in the coupling conditions
with the original partition function. We have
\a
{\d \over \d_{\al, r}} \ln Z_N(t,c) = {\rm Tr} \Big(Q^r(\al)\Big), \quad\quad
\al = 1,2 \label{ddZ}
\b
It is evident that, by using the equations (\ref{CC12},\ref{CC22}) above
we can express all the derivatives of $Z_N$ in terms of the elements of the
$Q$ matrices. For example
\a
{\d^2\over{\d t_{1,1}\d t_{\al,r}}}
\ln Z_N(t,c)=\Bigl(Q^r(\al)\Bigl)_{N,N-1},\qquad \al = 1,2\label{parti3}
\b
and so on.
We recall that the derivatives of $F(N,t,c) = \ln Z_N(t,c)$ are nothing but the
correlation functions of the model.

We can summarize the content of this section in the following

{\bf Proposition 2.1.} {\it The correlators (\ref{CF}) can be expressed
in terms of the entries of the matrices $Q(1)$ and $Q(2)$
via eq.(\ref{ddZ}) and
the like. In turn, these matrices must satisfy the coupling
conditions (\ref{coupling}) and the consistency conditions
(\ref{CC11}--\ref{CC22}).}

Knowing all the derivatives with respect
to the coupling parameters
we can reconstruct the partition function up to an overall integration
constant (depending only on $N$). The reconstructed free energy $F$ will
be a power series in the external couplings.

This theorem was proven in \cite{BX1}. It is a rigorous result when,
for example, highest negative even couplings guarantee
that the measure in (\ref{orth1}) is square--integrable and decreases
more then polynomially at infinity.
But for generic values of the couplings the above derivations are merely
heuristic.

However we notice that the consistency and coupling conditions
make sense for any value of the couplings, and also when the couplings are
infinite in number. In the latter case eqs.(\ref{CC12})
and (\ref{CC22}) form nothing but a very well--known
discrete integrable hierarchy, the Toda lattice hierarchy, \cite{UT}
(see also \cite{ANP}).
{}From these
considerations it is clearly very convenient to refer to the integrable
system formulation rather then to the original path integral formulation of our
problem. This allows us not only to extend our problem to a larger
region of the parameter space, but also to make full use of integrability.
Therefore we shift from the original problem to the new formulation:

{\it We call (unconstrained) two--matrix models all
the models
obtained by specifying a partition of the couplings between internal and
external couplings. Each such model is based on the Toda lattice hierarchy
and characterized by specific coupling conditions.}

{\bf Statement of the problem}. {\it Solve the integrable Toda lattice
hierarchy subject to the coupling conditions specific of a particular
unconstrained model and compute the correlators
as functions of the internal couplings via the relations
\a
< \tau_r> = {\rm Tr} \Big(Q(1)^r\Big), \quad\quad < \sigma_s >=
{\rm Tr} \Big(Q(2)^s\Big) \label{ddZ'}
\b
and the like.}

Once all the correlators are known, one can reconstruct
\footnote{Up to a constant depending only on $N$. There is a way
to determine this constant too, see \cite{BX2}, but we will not dwell upon
this point here.} the free energy $F$ by means of
\a
{\d \over \d t_{1, r}} F(N,t,c) = <\tau_r>,\qquad
{\d \over \d t_{2, r}} F(N,t,c) = < \sigma_r>
\b
$F$ will be a formal power series in the infinite set of external couplings.

Henceforth this will be the setup we refer to.

To end this subsection, we collect a few formulas we will need later on.
First, we will be using the following coordinatization of
the Jacobi matrices
\a
Q(1)=I_++\sum_i \sum_{l=0}^{m_1} a_l(i)E_{i,i-l}, \qquad\qquad\qquad
\q(2)=I_++\sum_i \sum_{l=0}^{m_2} b_l(i)E_{i,i-l}\label{jacobi1}
\b
One can immediately see that
\a
\Bigl(Q_+(1)\Bigl)_{ij}=\delta_{j,i+1}+a_0(i)\delta_{i,j},\qquad
\Bigl(Q_-(2)\Bigl)_{ij}=R(i)\delta_{j,i-1}\label{jacobi2}
\b
where $R({i+1}) \equiv h_{i+1}/h_i$.
As a consequence of this coordinatization, eq.(\ref{parti3}) gives in
particular the two important relations
\a
{\d^2\over{\d t^2_{1,1}}}
F(N,t,c)=a_1(N),\label{Fa1}
\b
Finally we write down explicitly
the $t_{1,1}$-- and $t_{2,1}$--flows, which will play a very important role
in what follows
\ai
&&{\partial\over{\partial t_{1,1}}}a_l(j)=a_{l+1}(j+1)-a_{l+1}(j)
+a_l(j)\Big(a_0(j)-a_0(j-l)\Big)\label{f11}\\
&&{\partial\over{\partial t_{2,1}}}a_l(j)=R({j-l+1})a_{l-1}(j)-R(j)a_{l-1}(j-1)
\label{f21}\\
&&{\partial\over{\partial t_{1,1}}}b_l(j)=R({j-l+1})b_{l-1}(j)-R(j)b_{l-1}(j-1)
\label{f11'}\\
&&{\partial\over{\partial t_{2,1}}}b_l(j)=b_{l+1}(j+1)-b_{l+1}(j)
+b_l(j)\Big(b_0(j)-b_0(j-l)\Big)\label{f21'}
\bj

\subsection{$W_{\infty}$ constraints.}

To solve the above stated problem we have to solve the flow equations of the
Toda lattice hierarchy subject to the coupling conditions (\ref{coupling}).
There is a way to put together flow equations and coupling conditions that
lead to an elegant algebraic structure, the $W$ constraints:
\vskip.2cm
{\bf Proposition 2.2}~~{\it The partition function of the unconstrained
two--matrix models satisfies the following $W$ constraints:
\a
W^{[r]}_n Z_N(t,c)=0, \quad\quad\quad
\tilde W^{[r]}_n Z_N(t,c)=0\quad r\geq0;~~n\geq-r,\label{Wc}
\b
where
\ai
W^{[r]}_n&\equiv& (-c)^n{\cal L}^{[r]}_n(1)-{\cal
L}^{[r+n]}_{-n}(2)\label{Wa}\\
\tilde W^{[r]}_n&\equiv& (-c)^n{\cal L}^{[r]}_n(2)-{\cal L}^{[r+n]}_{-n}(1)
\label{Wb}
\bj}

The generators ${\cal L}^{[r]}_n(1)$ are differential operators involving
$N$ and $t_{1,k}$, while ${\cal L}^{[r]}_n(2)$ have the same form
with $t_{1,k}$ replaced by $t_{2,k}$. One of the remarkable
aspects of (\ref{Wc}) is that the dependence on the coupling $c$ is nicely
factorized.
The ${\cal L}^{[r]}_n(1)$ satisfy the following $W_{\infty}$ algebra

\ai
&&\relax[{\cal L}^{[1]}_n(1), {\cal L}^{[1]}_m(1)]
 =(n-m){\cal L}^{[1]}_{n+m}(1)\label{LLa}\\
&&\relax[{\cal L}^{[2]}_n(1), {\cal L}^{[1]}_m(1)]
 =(n-2m){\cal L}^{[2]}_{n+m}(1)
 +m(m+1){\cal L}^{[1]}_{n+m}(1)\label{LLb}\\
&&\relax[{\cal L}^{[2]}_n(1),{\cal L}^{[2]}_m(1)]=
 2(n-m){\cal L}^{[3]}_{n+m}(1)-(n-m)(n+m+3){\cal L}^{[2]}_{n+m}(1)
 \label{LLc}
\bj
and in general
\a
[{\cal L}^{[r]}_n(1), {\cal L}^{[s]}_m(1)]=(sn-rm)
 {\cal L}^{[r+s-1]}_{n+m}(1)+\ldots,\label{LLgen}
\b
for $r,s\geq 1;~n\geq-r,m\geq-s$. Here dots denote lower than $r+s-1$ rank
operators. We notice that this $W_{\infty}$ algebra is not simple,
as it contains a Virasoro subalgebra spanned by the ${\cal L}^{[1]}_n(1)$'s.
For this reason, it is often called $W_{1+\infty}$ algebra. We also see
that once we know these generators and ${\cal L}^{[2]}_{-2}(1)$,
the remaining ones are
produced by the algebra itself.

The algebra of the ${\cal L}^{[r]}_n(2)$ is just a copy of the above one,
and the algebra satisfied by the $W^{[r]}_n$ and by $\tilde W^{[r]}_n$
is isomorphic to both.

The derivation of the $W$ constraints is very simple \cite{BX1}.
It consists of taking the coupling conditions (\ref{coupling}), multiplying
them by powers of $Q(1)$ and $Q(2)$, taking the finite trace
and using the flow equations of the Toda lattice hierarchy. This was
done in detail in ref.\cite{BX1}. There one can also find explicit expressions
of the generators, see also \cite{BX7}.

\subsection{Homogeneity and genus expansion.}

The CF's we compute are genus expanded. The genus expansion is strictly
connected with the homogeneity properties of the CF's.
As we will see the contribution
pertinent to any genus is a homogeneous function of the couplings (and $N$)
with respect to appropriate degrees assigned to all the involved quantities.
Precisely, we assign to the couplings the following degrees
\a
deg(~) \equiv [~],\qquad [t_{\al,k}] = y - y_\al k,\qquad [N]=y,
\qquad [c] = y -y_1 - y_2\label{degree}
\b
where $y, y_1, y_2$ are arbitrary constants. Here and in the
following $N$ is treated as a coupling $t_{1,0}=t_{2,0}$.

If we rescale the couplings as follows
\a
t_{\al,k} \rightarrow \lambda^{[t_{\al,k}]} t_{\al,k}\0
\b
on the basis of the analysis of Bessis, Itzykson and Zuber, \cite{BIZ}, we
expect the free energy to scale like
\a
F \rightarrow \sum_{h=0}^\infty \lambda^{2y(1-h)} F^{(h)}\label{genusexp}
\b
where $F^{(h)}$ is the genus h contribution. In other words
\a
\relax [F^{(h)}] = 2y(1-h) \label{degFh}
\b
The CF's will be expanded accordingly.
Such expectation, based on a path integral analysis, remains true in our setup
due to the fact that the homogeneity properties carry over to the Toda lattice
hierarchy. To this end we have simply to consider a genus expansion for all
the coordinate fields that appear in $Q(1)$ and $Q(2)$, see (\ref{jacobi1},
\ref{jacobi2}).
The Toda lattice hierarchy splits accordingly.
In genus 0 the following assignments
\a
\relax [a_l^{(0)}] = (l+1) y_1,\qquad [b_l^{(0)}] = (l+1) y_2,\qquad
[R^{(0)}] = y_1 + y_2 \label{degToda}
\b
correspond exactly to the assignments (\ref{degree}) and $[F^{(0)}] = 2y$.

It is very common to replace the matrix size $N$ with a continuum variable,
say $x$. This is permitted provided one rescales all the quantities involved
according to the above degrees, \cite{BX2}.

\section{Correlation functions in (unconstrained) matrix models.}

\setcounter{equation}{0}
\setcounter{footnote}{0}

We have (at least) three methods to calculate CF's. The first consists
of directly solving
the $W$ constraints; the second consists of determining from the
coupling conditions the explicit form of $Q(1)$ and $Q(2)$ and then using
the flows of the discrete Toda lattice hierarchy; the third method is
based on passing from the discrete hierarchy to a purely differential one
and integrating the flows of the latter. The first method has been shown
in a number of examples, \cite{BX2} and \cite{BX7}. Moreover we will see it
at work in the constrained models. Therefore we skip it here and pass
directly to the second method.

\subsection{Solving the coupling conditions: ${\cal M}_{2,2}$ model}.

This method is based on explicitly
solving the coupling constraints (\ref{coupling}). It is then elementary
to compute correlators by means of eq.(\ref{ddZ}) and the lattice Toda
flow equations. First we discuss in detail the model
${\cal M}_{2,2}$, i.e. the purely Gaussian case.
For simplicity we set, in the following,
\a
t_{1,k} \equiv t_k,\qquad t_{2,k}\equiv s_k\0
\b

{\bf Lemma 3.1} {\it  The matrices $Q(1)$ and $Q(2)$ relevant for the model
${\cal M}_{2,2}$ are specified, in reference to the coordinatization
(\ref{jacobi1},\ref{jacobi2}),  by the following coordinates
\a
&&a_0(n) \equiv a_0= \frac{c s_1 - 2 s_2 t_1}{4s_2t_2-c^2},\qquad
a_1(n) = -\frac{ 2 s_2 n}{4s_2t_2-c^2}, \label{Qm22}\\
&&b_0(n)\equiv b_0 = \frac{c t_1 - 2 t_2 s_1}{4s_2t_2-c^2},\qquad
b_1(n)= -\frac{2t_2n}{4s_2t_2-c^2},\qquad
R(n) = \frac{nc}{4s_2t_2-c^2}\0
\b
The remaining coordinates vanish.}

Proof. The coupling conditions (\ref{coupling}) for the model
${\cal M}_{2,2}$ are
\a
&&P^\circ (1) +t_1 + 2t_2 Q(1) + c Q(2) =0\0\\
&& {\overline {\cal P}}^\circ (2) + s_1 + 2s_2 Q(2) + c Q(1) =0\0
\b
Using the fact that $P^\circ_{n,n-1} (i) =n$ for $i=1,2$, they can be
explicitly written in terms of the coordinates as follows.
\a
2t_2R(n) + c b_1(n) =0  && c a_1(n) + 2 s_2 R(n) =0\0\\
t_1 + 2t_2 a_0(n) + c b_0(n) = 0 && s_1 + c a_0(n) + 2s_2 b_0(n) =0\0\\
n + 2t_2 a_1(n) + c R(n) =0 && cR(n) + 2 s_2 b_1(n) +n =0\0
\b
These equations can be easily solved and give (\ref{Qm22}).

{\bf Proposition 3.2} {\it The exact one--point and two--point correlators of
the model ${\cal M}_{2,2}$  are given by the following formulas
\a
<\tau_r> &=& \sum_{2l=0}^r \sum_{k=0}^l \frac {r!2^{-k}}{(r-2l)! k!(l-k)!}
\left(\bac N \\ l-k+1\ea\right) \Big(\frac {2 s_2}{c^2-4s_2t_2 }\Big)^{l-k}
\Big( \frac{2 s_2 t_1-cs_1}{c^2-4s_2t_2}\Big)^{r-2l} \label{taun}
\b}

Proof. To start with it is convenient to rewrite $Q(1)$ and $Q(2)$ as
\a
Q(i) = \alpha_i I_+ + \beta_i I + \gamma_i \epsilon_- ,
\qquad i= 1,2 \label{Qepsilon}
\b
where
\a
I_+ = \sum_{n=0}^\infty E_{n,n+1}, \qquad
I = \sum_{i=0}^\infty E_{n,n},\qquad \epsilon_- = \sum_{n=1}^\infty
nE_{n,n-1}\0
\b
and
\a
\al_1 =1, \quad \beta_1 = a_0, \quad \gamma_1 = \frac {2 s_2}{c^2-4s_2t_2 },
\quad
\al_2 = - \frac{2t_2}{c}, \quad \beta_2 = b_0,\quad \gamma_2 = \frac{c}
{4s_2t_2-c^2 }\label{symbols}
\b
By means of the formulas
\a
\relax [I_+ , \epsilon_-] = I ,\qquad {\rm Tr} \Big(\epsilon_-^k
I_+^l \Big) = \delta_{k,l} \sum_{n=k}^{N-1} \frac {n!}{(n-k)!} =
\delta_{n,k}~ k! \left( \bac N\\ k+1\ea\right) \label{formulas}
\b
we can now make explicit computations. For example
\a
{\rm Tr} \Big( Q(1)^r\Big) &=& \sum_{2l =0}^r \left( \bac r\\2l\ea\right)
{\rm Tr} (I_+ + \gamma_1\epsilon_-)^{2l} \beta_1^{r-2l} \0\\
&=& \sum_{2l=0}^r \sum_{k=0}^l \frac {r!2^{-k}}{(r-2l)! k!(l-k)!}
\left(\bac N \\ l-k+1\ea\right) \gamma_1^{l} \beta_1^{r-2l} \label{Q1n}
\b
{}From this formula, using (\ref{ddZ}), we can immediately get eq.(\ref{taun})
above. In a similar way we can derive $<\sigma_r>$.

Finally using the genus expansion (\ref{genusexp}) we can extract the
genus by genus correlators.

{\bf Corollary 3.1} {\it The genus h contributions to the one-- and two--point
CF's in the model ${\cal M}_{2,2}$ are
\a
<\tau_r>_h = \sum_{2l=0}^r \sum_{k=0}^l \frac {(-1)^{2h-k}2^{-k}r!
\beta_{2h-k}(l-k)N^{l+1-2h}}{(r-2l)! k!(l-k)!(l-k+1)!}
 \Big(\frac {2 s_2}{c^2-4s_2t_2 }\Big)^{l}
\Big( \frac{2 s_2 t_1-cs_1}{c^2-4s_2t_2}\Big)^{r-2l} \label{taunh}
\b
where
\a
\beta_k(r) = \sum_{1\leq r_1 \leq r_2 \ldots \leq r_k \leq r}
r_1r_2\ldots r_k,\qquad 1\leq k\leq r,\qquad \beta_0(r) =1,
\qquad\beta_k(r) = 0 \quad {\rm otherwise}\0
\b}

Due to the $\beta$ factor the sum over $l$ in (\ref{taunh}) starts at $2h$
and the sum over $k$ ends at $2h$.

For the two--point correlators, see Appendix.

We have given the above proof in some detail since it constitutes a model
for all the other more complicated cases.
In fact there is nothing new when we consider the ${\cal M}_{p,1}$
models. They can be solved exactly in the same way.
New features appear in the case of the ${\cal M}_{p_1,p_2}$ models
with $p_1,p_2> 1$ and $p_1+p_2 >4$, since the constraints give rise to
non--linear algebraic relations for the coordinates. Let us see the simplest
possible example of this situation.

\subsection{Solving the coupling conditions: ${\cal M}_{3,2}$ model}

The coupling conditions of the ${\cal M}_{3,2}$ model are
\a
&& P^\circ(1) + 3t_3 Q(1)^2 + 2t_2 Q(1)  + t_1 + c Q(1) =0\0\\
&& \bar {\cal P}^\circ_2 + 2s_2 Q(2) + s_1 + c Q(1) =0\label{coum32}
\b
Using the coordinatization (\ref{jacobi1}) and (\ref{jacobi2})
we find that the fields $a_l(n), b_l(n), R(n)$  must satisfy the equations
\a
&& cb_2(n) + 3t_3 R(n) R(n-1) =0 \0\\
&& 2t_2 R(n) + c b_1(n) + 3t_3 R(n)\Big (a_0(n) + a_0(n-1)\Big) =0\0\\
&& 3t_3 \Big( a_0(n)^2 + a_1(n) + a_1(n+1) \Big)
+ 2t_2 a_0(n) +t_1 + c b_0(n) =0\0\\
&& n+ 3t_3 a_1(n) \Big( a_0(n) + a_0(n-1)\Big) + 2t_2 a_1(n) + c R(n) =0\0\\
&&2s_2 R(n) +ca_1(n)=0\label{coum32'}\\
&& 2s_2 b_0(n) + s_1 + c a_0(n) =0 \0\\
&& n + 2 s_2 b_1(n) + c R(n) =0 \0
\b
One easily realizes that the second, fourth,
fifth and seventh equations are linearly dependent.
Finally one has
\a
&&a_1(n) = - \frac {2s_2}{c} R(n), \qquad
b_0(n) = - \frac{s_1 + c a_0(n)}{2 s_2}\0\\
&& b_1(n) = - \frac{n + c R(n) }{2 s_2}, \qquad  b_2(n) =
- \frac{3t_3}{c} R(n)R(n-1)\0
\b
and the recursion  relations
\a
&& a_0(n) + a_0(n-1) = -\frac{2t_2}{3t_3}
+ \frac {c}{6s_2t_3} \Big( c + \frac{n}{R(n)}\Big)\label{recur1}\\
&& R(n+1)+ R(n) = \frac{c}{2s_2} a_0(n)^2
+ \Big( \frac{2 c t_2}{6s_2t_3}
- \frac{c^3}{12 s_2^2 t_3}\Big) a_0(n)
-\frac{c^2 s_1}{12 s_2^2t_3} + \frac {ct_1}{6s_2t_3} \label{recur2}
\b
These recursion relations can be solved exactly in complete generality,
although the final formulas may look very cumbersome. However, for our present
purposes, it will be sufficient to see the solutions in genus 0.
In genus 0 the above equations become:
\a
&&a_1(n) = - \frac {2s_2}{c} R(n), \qquad
b_0(n) = - \frac{s_1 + c a_0(n)}{2 s_2}\0\\
&& b_1(n) = - \frac{n + c R(n) }{2 s_2}, \qquad
b_2(n) = - \frac{3t_3}{c} R(n)^2\label{couplm32}
\b
and the recursion  relations
\a
&& 2a_0(n) = -\frac{2t_2}{3t_3} + \frac {c}{6s_2t_3}
\Big( c + \frac{n}{R(n)}\Big)\label{recur1'}\\
&&2R(n) = \frac{c}{2s_2} a_0(n)^2 + \Big( \frac{2 c t_2}
{6s_2t_3} - \frac{c^3}{12 s_2^2 t_3}\Big) a_0(n)
-\frac{c^2 s_1}{12 s_2^2t_3} + \frac {ct_1}{6s_2t_3} \label{recur2'}
\b
This leads to a cubic equations for $a_0$. Once this equation is solved
with the standard formulae, all the fields are completely determined.
Since they are not particularly illuminating, we do not write down here
the explicit solutions. We notice however that, once we know them, it is
possible to write down immediately an integral expression for the correlation
functions. For example, using the same formulas as in the previous subsection,
one gets
\a
<\tau_r>_0 = \sum_{l\geq r/2}^r \frac {r!}{((r-l)!)^2 (2l-r)!}\int^xdn
\bar a_0^{2l-r}\bar a_1^{r-l}\label{taurm32}
\b
where $\bar a_0,\bar a_1$ are the solutions of the above algebraic system,
and we have promoted $n$ and $N$ to continuum variables and called
the latter $x$. In a similar way we can obtain all the correlators we wish.

As we see from this example, the method is the same as in the ${\cal M}_{2,2}$
model, the only additional difficulty being the solution of a third order
algebraic equation. When we come to more complicated ${\cal M}_{p_1,p_2}$
models, the method remain the same but we are faced with the problem of solving
higher order algebraic equations.

It has been shown in \cite{BX2} (see also \cite{BX8})
that the model ${\cal M}_{0,0}$ represents the $c=1$ string theory at the
self--dual point. Any ${\cal M}_{p_1,p_2}$ model represents the perturbation
of the former by the corresponding tachyonic states. We have shown that
these perturbations can, at least in principle, be solved. However we do not
see
any point in pushing the analysis further in this direction. We would rather
like to concentrate from now on on a related interesting problem: {\it can
one obtain from non--Gaussian ${\cal M}_{p_1,p_2}$ models `simple' submodels,
in the sense that, for example, the correlators are polynomials of the
couplings?} The answer is yes, and the submodels are obtained by imposing
constraints in the coordinates of the ${\cal M}_{p_1,p_2}$ models.
The submodels are called {\it constrained two--matrix models} and to the
analysis of some of them are devoted the next three sections.

\section{Differential Hierarchies of Two--Matrix Models.}

\setcounter{equation}{0}
\setcounter{footnote}{0}

One possible characterization of the constrained models is by means
of the differential integrable hierarchies they correspond to.

Let us return to section 2.
We saw there that two--matrix models can be represented by means of
coupled discrete linear systems, whose consistency conditions give rise to
the Toda lattice hierarchy. Here we review the method, used in \cite{BX1},
to transform the discrete linear
systems into equivalent differential systems whose consistency conditions are
purely differential hierarchies. This is tantamount to separating
the $N$ dependence from the dependence on the couplings.
This section is introduced for the sake of completeness: we collect and
try to render as plausible as possible the
results obtained in \cite{BX1},\cite{BX4} we will need in the following.

The clue to the construction are the first flows, i.e.
the $t_{1,1}$ and $t_{2,1}$ flows. For the sake of simplicity let us
consider the {\it system I} and the flow (\ref{f11}). Let us consider
the generic situation in which $Q(1)$ has $m_1 = p_2-1$ lower diagonal lines
(see the parametrization (\ref{jacobi1})). To begin with we notice that
\a
{\partial\over{\partial t_{1,1}}}\Psi_n=\Psi_{n+1}+a_0(n)\Psi_n\label{t11psi}
\b
and adopt for any function $f(t)$ the convention
$f'\equiv {{\partial f}\over{\partial t_{1,1}}}\equiv \partial f $.
We can rewrite
\a
\Psi_{n}=\hat B_{n}\Psi_{n+1}\label{AB}
\b
where
\a
{\hat B}_n\equiv {1\over{\partial-a_0(n)}}=\d^{-1}\sum_{l=0}^{\infty}
\bigl(a_0(n)\d^{-1}\bigl)^l\label{Bn}
\b
In so doing we implicitly understand that the framework in which we
operate is that of the pseudodifferential calculus, see for example
\cite{Dickey}.

It is now an easy exercise to prove that the discrete spectral
equation
\a
Q(1)\Psi(\lambda_1)=\lambda_1\Psi(\lambda_1)\0
\b
is transformed into the pseudodifferential one
\a
L_n(1)\Psi_n=\lambda_1\Psi_n\label{spetral1}
\b
where
\a
&&L_n(1)=\d+\sum_{l=1}^{m_1}a_l(n)\hat B_{n-l}\hat B_{n-l+1}\ldots
\hat B_{n-1}\label{Ln}\\
&&\qquad=\d+\sum_{l=1}^{m_1}a_l(n)
{1\over{\d-a_0(n-l)}}\cdot
{1\over{\d-a_0(n-l+1)}}\cdots{1\over{\d-a_0(n-1)}}\0
\b

Proceeding in the same way for the other equations of {\it system I}
we obtain the new system in differential form
\a
\left\{\ba{ll} L_n(1)\Psi_n=\lambda_1\Psi_n\\\noal
\frac {\d}{ \d t_{1,r}}\Psi_n=\Bigl(L^r_n(1)\Bigl)_+\Psi_n,
\ea\right.\label{diffls1}
\b
The subscript + appended to a pseudo--differential operator
represents the purely
differential part of it. The subscript -- represents the complementary part.

Let us come now to the $n$ dependence of the above equations.
The operator $L_n(1)$ in (\ref{Ln}) depends on the coordinates
of different lattice points. To deal with this complication,
we introduce $m_1$ ``fields'' $S_1, ... , S_{m_1}$, related
to the ``field'' $a_0$ in the following way
\a
S_i(n) \equiv a_0 (n-i) \label{Sa0}
\b
Then we can rewrite $L_n(1)$ in the following way
\a
L_n(1)=\d+\sum_{l=1}^{m_1}a_l(n)
{1\over{\d-S_{l}(n)}}\cdot
{1\over{\d-S_{l-1}(n)}}\cdots{1\over{\d-S_1(n)}}\label{Ln1}
\b
with the result that $L_n(1)$ is expressed in terms of fields
evaluated at the same lattice point. Of course the fields
$S_i$ are not independent.
However we will consider these fields as completely independent from
one another in all the intermediate steps of our calculations and only
eventually impose the condition (\ref{Sa0}).

To further simplify the notation we will consider henceforth the lattice label
$n$ on the same footing as the couplings and write
\a
a_i(n, ...) \equiv a_i (n) ( ...), \quad\quad \quad
S_i(n, ...) \equiv S_i (n)( ...) \0
\b
where the dots denote the dependence on $t_{1,k}, t_{2,k}$ and  $g$.
So the expression of $L_1(n)$ gets further simplified to
\a
L=\d+\sum_{l=1}^{m}a_l
{1\over{\d-S_{l}}}\cdot
{1\over{\d-S_{l-1}}}\cdots{1\over{\d-S_1}}\label{defL}
\b
where, for simplicity, we have dropped the label $(1)$ too.
This simplified form is the one we constantly
refer to throughout this and the following section.

A similar treatment can be applied to the second linear system as well.
Therefore {\it the information concerning matrix models can be stored in
two differential linear systems + the first flow equations (\ref{f11},
\ref{f21},\ref{f11'},\ref{f21'})}. The former determine the dependence
on the couplings, while the latter fix the dependence on $N$.
Therefore what we have accomplished so far is the separation of the
dependence on $N$ from the dependence on the couplings.

{}From now on we refer
to the consistency conditions
\a
\frac{\d}{\d t_r} L= [ (L^r)_+, L] \label{CC}
\b
where $t_r\equiv t_{1,r}$. (\ref{CC}) are integrable hierarchies,
\cite{BX5}\cite{BX6}, which are classified by the number $2m$ of fields. The
pseudodifferential operator $L$ in (\ref{defL}) is
the relevant Lax operator.

We can easily locate these hierarchies in a well--known framework.
In fact $L$ is nothing but a particular realization of the KP operator
\a
L_{KP}= \d+\sum_{l=1}^{\infty}w_l\d^{-l}\0
\b
In general, $w_l$ are unrestricted coordinates, while in the realizations
(\ref{defL}) they are precise functions of the fields $a_l$ and $S_l$ and their
derivatives. But that is not all,
for one can obtain new integrable hierarchies {\it via hamiltonian reduction}.
Each integrable hierarchy characterizes a different model.
In the case of a reduced hierarchy, we call the
corresponding model {\it a reduced model}. These reduced models will be shown
to essentially coincide with the constrained ones.

The results can be synthesized as follows.

{\bf Summary.} Starting from the Lax operator (\ref{defL}) with given $m$ one
can show that:

\noindent 1) {\it there are $m+1$ distinct differential integrable
hierarchies which
are obtained by suppressing successively the fields $S_l$ with the Dirac
procedure;

\noindent 2) of each such hierarchy it is possible to write down the
relevant Lax operator;

\noindent 3) at the end of this cascade procedure we find the $(m+1)$--th
KdV hierarchy}.

Therefore {\it for every $p=m+1$ we have $p$ systems or hierarchies,
denoted henceforth with the symbol
${\cal S}_p^l$, where $l$ counts the number of nonvanishing $S$ fields,
$0\leq l\leq m$. In
particular the case $l=m$ corresponds to the $2m$--field representation
of the KP hierarchy, while $l=0$ corresponds to the $p$--KdV hierarchy.}

The above is general and holds
for the more complex systems with $m>2$. The general case was treated in
\cite{BX5} (see also \cite{BX4}).

\subsection{Examples: the KdV and Boussinesq hierarchies}

The simplest example of $L$, (\ref{defL}), corresponds to $m=1$.
It gives rise to the NLS hierarchy.
\a
L=\d +a_1\frac{1}{\d-S_1}\label{NLS}
\b
If we impose the constraint $S_1=0$, the second Poisson structure can be
reduced via the Dirac procedure and leads to the a classical version of the
Virasoro algebra. The corresponding integrable hierarchy is the KdV hierarchy.
Later on we will need the recursion relations for the flows of this hierarchy.
They are introduced as follows. Let
\a
F_{2k}(x) = \frac {\delta H_{2k}}{\delta a(x)}\label{F2k}
\b
where $H_{2k}$ are the Hamiltonians, whose explicit form can be derived from
the Lax operator, \cite{BX4},\cite{BX5}, and $a\equiv a_1$.
Then, imposing the compatibility between the two Poisson brackets,
characteristic of the hierarchy,
we find the recursion relation for the flows
\a
\frac{\d a}{\d t_{2k+1}} = F_{2k+2}' =D_F F_{2k}, \qquad
D_F=\d^3 + 4a \d + 2a' \label{recurKdV}
\b
with $F_2=a$.

The simplest integrable system that appears in matrix models
after the NLS system is the four--field representation of the KP hierarchy
($m=2$). It naturally leads, via reduction, to the Boussinesq hierarchy.
Let us describe the latter as concisely as possible. The Boussinesq system is
described by two fields $a_1$ and $a_2$.
The Lax operator is
\a
 L_B= \d^3 + a_1 \d +a_2\label{Boussop}
\b
The second Poisson structure is nothing but the classical $W_3$ algebra.
The second flow equations are
\a
\ddt2 a_1=2a_2'-a_1'',\quad\quad\quad
\ddt2 a_2=a_2^{''}-\frac{2}{3}(a_1a_1'+a_1''')\label{Bouss}
\b
This is known as the Boussinesq equation (in parametric form) and it is the
first of an integrable hierarchy of equations (the Boussinesq or
3--KdV or ${\cal S}_3^0$ hierarchy).

Like in the KdV case, we give the recursion relations that allow us to
calculate all the flows. Let us define
\a
F_r (x)= \frac{\delta H_r}{\delta a_1(x)}, \qquad
G_r (x) = \frac{\delta H_r}{\delta a_2(x)}, \qquad r \neq 3n\0
\b
Then imposing the compatibility of the two relevant Poisson brackets,
we find the following recursion relations
\a
\frac{\d a_1}{\d t_r} = 3G_{r+3}'
&=& D_{GG} G_r + D_{GF}F_r\label{recurGr}\\
\frac{\d a_2}{\d t_r} = 3F_{r+3}'
&=& D_{FG}G_r + D_{FF}F_r\label{recurFr}
\b
with $F_1=1, ~G_1 =0$ and $F_2=0,~G_2=1$.
The differential operators are
\ai
&&D_{GG} = 3a_2 \d +2a_2' - a_1 \d^2 - 2 a_1' \d - a_1'' - \d^4\label{DGG}\\
&&D_{GF} = 2\d^3 + 2 a_1 \d + a_1'\label{DGF}\\
&&D_{FG} = 2a_2'\d + a_2'' - {2\over 3}\Big(a_1 a_1' + a_1^2\d + 2 a_1\d^3
+ a_1''' + 3a_1''\d + 3 a_1'\d^2 + \d^5\Big)\label{DFG}\\
&&D_{FF} = \d^4 + a_1 \d^2 + a_2' + 3a_2\d \label{DFF}
\bj

\section{Correlators in reduced models}

In this section we show that, starting from the p--KdV or ${\cal S}_p^0$
hierarchy and borrowing some information from matrix models, we can
define models, i.e. we can define (and compute) a full set
of correlators -- which turn out to essentially coincide with the
constrained models we will meet later on. Since however the construction
in this section is somewhat heuristic, and, in particular, it does not permit
us to carefully fix all the normalizations, we prefer to distinguish these
models from the constrained two--matrix models of the following section:
as they are based on reduced hierarchies, we refer to them as {\it reduced
models}. We call $\widehat{\cal M}_p$ the reduced model based on the p--KdV
hierarchy.
In presenting the reduced models before the constrained two--matrix models,
which are the true objectives of our research, we follow a historical and,
hopefully, pedagogical order.

The essential definition of the reduced models is as follows: we define the
correlators by identifying the field $a_1$ with the two--point function
$<\tau_1\tau_1>$, i.e. we borrow from the matrix models eq.(\ref{Fa1}), and
differentiate (or integrate) $a_1$ with respect to the couplings, as necessary.
Moreover we only consider the dependence on the
$t_{k}\equiv t_{1,k}$ couplings and disregard the remaining ones.
At this point the flows of the relevant hierarchy allow us to calculate
the correlators, at least up to some constants -- we are going
to see some examples later on. The reason for this is as follows.
A part of the information concerning
the coupling conditions is in fact stored in the differential system of
the model: the Lax operator inherits the information contained in the second
equation (\ref{coupling}) via the number of non--vanishing
diagonal lines of the original $Q(1)$ matrix. Therefore it is not surprising
that the flow equations are almost enough to determine the CF's. However
not all the information concerning the coupling conditions is contained
in the differential hierarchy which characterizes the model, the first equation
(\ref{coupling}) is not, and this is reflected in the undetermined constants
that appear when we try to calculate the correlators.

Let us see this point in detail in an explicit example.

\subsection{The KdV hierarchy and the associated $\widehat{\cal M}_2$ model }

We showed in section 4 that we are allowed to impose the constraint $S=0$
on the NLS system while preserving integrability. In other words there
is a consistent subsystem of the NLS system, of which we can easily compute
the flows, (\ref{recurKdV}). These are the KdV flows.
We recall that only the odd flows survive the reduction.
Therefore the $t_{2n}$ are disregarded. It is therefore natural to forget
$t_0\equiv N$ as well.

To start with let us define the critical points for this model:

{\it The k--th critical point of the $\widehat{\cal M}_2$ model
are defined by $2(2k+1)t_{2k+1} =-1$ and $t_l=0$ for $l\neq 1, 2k+1$.}

For the origin of this terminology and further properties of critical
points in matrix models see \cite{BX7}. We will see next that the above
critical point corresponds to a two--matrix model with
a $V_1$ potential of order $2k+1$ and a $V_2$ potential of cubic order.
Let us notice 1) that the correlators of $\widehat{\cal M}_2$ at the various
critical points are functions of $t_1$ alone, 2) that in order to preserve the
homogeneity properties at the $k$--th critical point we have to set
$y= y_1(2k+1)$ in (\ref{degree}).

In the following we study the first critical point, $k=1$. On the basis of
the assignments of section 2.3 we have $[a]=[t_1] = 2y_1$, in genus 0.
Therefore it must be: $a\sim t_1$. The proportionality constant
can be absorbed with a rescaling (provided it is non--vanishing, which is
the case as we shall see). So we start from the position
\a
a = t_1\label{ansa}
\b
Then we have:

{\bf Lemma 5.1} {\it As a consequence of (\ref{ansa}), the functions $F_{2n}$
relevant to the KdV hierarchy are given by
\a
F_{2n} = \sum_{h=0}^\infty a(n,h) t^{n-3h}, \qquad n\geq 3h\label{F2n}
\b
where
\a
a(n,h) = \frac{2^{n-1}}{12^h h!} \frac {(2n-1)!!}{(n-3h)!}, \label{anh}
\b
and $r!!$ is the 2 by 2 factorial, i.e. $r!! = r(r-2)\ldots$ up to
$1~{\rm or}~2$.}

Proof. We insert the expression (\ref{F2n}) into (\ref{recurKdV}) and obtain
the recursion relation
\a
a(n+1,h) = (n-3h+2)(n-3h+3) a(n,h-1) + 2 \frac{2n-6h+1}{n-3h+1} a(n,h)
\label{recuranh}
\b
for the coefficients $a(n,h)$. One can immediately verify that
(\ref{anh}) is a solution of (\ref{recuranh}), but it is not unique. While
integrating (\ref{recuranh}) one has to specify what $b_h \equiv a(3h,h)$
are $\forall h$. The latter are the coefficients of $(a')^{2h}$ in $F_{6h}$
and satisfy the recurrence relation
\a
3hb_h = 2(6h-1)(6h-3)(6h-5) b_{h-1}, \qquad b_1=5\0
\b
One immediately gets (\ref{anh}). This ends the proof of the Lemma.

{\bf Proposition 5.1} {\it The exact one--point correlators of the
$\widehat{\cal M}_2$ model at the first critical point are
\a
<\tau_{2n-1}> = \sum_{h=0}^\infty <\tau_{2n-1}>_h=
\sum_{h=0}^\infty \frac{2^{n-1}}{12^h h!} \frac {(2n-1)!!}
{(n-3h+1)!} t^{n-3h+1},\qquad n\geq 3h-1\label{1pkdv}
\b
where the genus expansion is explicitly exhibited.}

Proof (partial). We have simply to recall that
$F_{2n}= <\tau_{2n-1}\tau_1>$ and integrate
over $t_1$. We obtain (\ref{1pkdv}) for $n\geq 3h$. The values of
$<\tau_{6h-3}>_h$ (i.e. $n= 3h-1$), which are obtained by simple extension of
this result, are also correct, but strictly speaking they do not follow from
the
previous argument: they are pure integration constants and cannot be
obtained from the
flows alone. We will be able to completely justify eq.(\ref{1pkdv}) only by
means of the W constraints. It is in fact the $W$--constraints that
completely determine such constants.

\subsection{$W$--constraints of the $\widehat{\cal M}_2$ model}

Some information concerning the coupling conditions is not contained in the
differential KdV hierarchy. In order to retrieve it we have to use the
$W$ constraints. The point is that they cannot be the same as in the unreduced
models, since the hierarchy underlying the model has changed, and we recall
once again that the $W$ constraints are based on the flow equations. Therefore
we have to reconstruct {\it effective} $W$ constraints on the basis of the
reduced hierarchy.
Let us argue as follows. In reduced models we are interested in
solutions that do not depend on the second sector (i.e. on $t_{2,k}$).
If we look at eq.(\ref{Wc}), we see that such solutions should therefore
satisfy
to $W$--constraints of the form
\a
{\cal L}^{[r]}_n(1)Z_N=0,\qquad r\geq1;\qquad n\geq-r\label{provv}
\b
Consequently we look for $W$ constraints of this type, with generators
belonging
to a $W$ algebra, which are however compatible
with the KdV hierarchy. It is easy to see that the {\it universal} generators
in (\ref{Wc}), (see \cite{BX1}), do not satisfy the KdV flows. We find instead

{\bf Proposition 5.2} {\it The effective $W$ constraints for
$\widehat{\cal M}_2$ take the form
\a
L_n \sqrt Z =0 , \qquad \qquad n\geq -1\label{WKdV}
\b
where
\a
&&L_{-1}= \sum_{k=1}^\infty (k+{1\over 2})t_{2k+1}
\frac {\partial}{\partial t_{2k-1}} +\frac {t_1^2}{16}\0\\
&&L_0=\sum_{k=0}^\infty (k+{1\over 2})t_{2k+1}
\frac {\partial}{\partial t_{2k+1}}+{1\over {16}}\label{VirLn}\\
&&L_n=\sum_{k=0}^\infty (k+{1\over 2})t_{2k+1}
\frac {\partial}{\partial t_{2k+2n+1}}+
\sum_{k=0}^{n-1} \frac {\partial^2}{\partial t_{2k+1}\partial t_{2n-2k-1}},
{}~~~~n>0\0
\b
These generators satisfy the commutation relations of the Virasoro algebra.}

Proof. Let us prove first that (\ref{WKdV}) are in agreement with
the KdV flows. To this end we differentiate (\ref{WKdV}) with $n>0$
with respect to $t_1$.
Using the definition of $F_{2k}$, we can write (remember the notations
introduced after eq.(4.1))
\a
&&\sum_{k=0}^\infty (2k+1)t_{2k+1} F_{2k+2n+2} +\d^{-1}F_{2n+2} \0\\
&&~~~~~~~~~+\sum_{k=0}^{n-1}
\Big(2 \frac{\d }{\d t_{2k+1}}F_{2n-2k} + F_{2k+2} \d^{-1}
F_{2n-2k} + \d^{-1} F_{2k+2}F_{2n-2k}\Big)=0\0
\b
Here $\d^{-1}$ is understood in the sense of the pseudodifferential
calculus and denotes indefinite integration (see below for further
specifications).
Now we apply to it the recursion operator $D_F$. What we obtain,
by using eq.(\ref{recurKdV}), is nothing but the constraint $L_{n+1}\sqrt Z=0$
differentiated twice with respect to $x$. To see this we have to apply
the remarkable formula
\a
F_{2n+4} &=& F_{2n+2}'' + 3F_2F_{2n+2} +
\sum_{k=0}^{n-1}\Big(2F_{2k+2}F_{2n-2k}''
-F_{2k+2}'F_{2n-2k}'\0\\
&& +4F_2F_{2k+2}F_{2n-2k} - F_{2k+2} F_{2n-2k+2}\Big)\0
\b
which can be obtained once again from the recursion relation (\ref{recurKdV}).
As for the cases $n=0$ and $n=-1$, which have not been included in
the above argument, they can be explicitly verified.

What we have done so far amounts to saying that starting from $L_{-1}\sqrt Z=0$
and successively applying the operator ${\cal O}= \d^{-2}D_F \d$, we obtain
all the $L_n\sqrt Z=0$. Here we have to exercise some care with
the double integration
$\d^{-2}$. $\d^{-1}$ represents an indefinite integration which preserves
the homogeneity properties. This is a perfectly well defined operation unless
the output of it has degree 0. In such a case a numerical integration constant
may appear. Now, in $Z^{-1/2}L_n Z^{1/2}$ there appear contributions of
degree $2y(1-h)+ 2ny_1$, with $n=0,1,2,\ldots$. So, since $y$ and $y_1$
are generic numbers, the only dangerous case (in the above sense) is when
$h=1, n=0$. In other words when we pass from $L_{-1} \sqrt Z=0$ to
$L_0 \sqrt Z=0$ by applying ${\cal O}$ we are not guaranteed that the
appropriate constant is given by the ${1\over 16}$ present in $L_0$. However
at this point we make the request that $[L_1,L_{-1}] = 2L_0$, and this
unambiguously fixes such constant. It remains for us to justify
$L_{-1}\sqrt Z =0$ (which is often referred to as the string equation) or
rather the term $\sim t_1^2$ in $L_{-1}$. From the degree analysis
one sees that the only possible polynomial of the couplings one can write
is $t_1^2$. Its coefficient is determined by the requirement that, applying
the recursion device to $L_{-1}$, one gets $L_0$.

Finally, we do not look for higher tensor constraints since eq.(\ref{WKdV})
is enough to determine everything.

On the basis of eq.(\ref{WKdV}) one can complete the proof of Proposition 5.1.

\subsection{The Boussinesq hierarchy and the associated
$\widehat{\cal M}_3$ model.}

The 3--KdV or Boussinesq hierarchy was introduced in 4.2 as a reduced
hierarchy. The corresponding model is denoted $\widehat{\cal M}_3$.
It is described by two fields $a_1$ and $a_2$ and is
specified by the Lax operator
\a
L=\d^3+a_1\d + a_2 \label{L03}
\b
In the Boussinesq hierarchy the $t_{3k}$ flows with $k = 0,1,2,3...$
do not appear.

The correlation function interpretation of the fields $a_1$ and $a_2$
is given by eq.(\ref{Fa1}) and the first of (\ref{Bouss}):
\a
a_1 = <\tau_1\tau_1 >,\quad\quad 2a_2 =<\tau_1\tau_2 >
+<\tau_1\tau_1\tau_1>,
\label{a1a2}
\b

Now we proceed as in the KdV case (but skip many details).
The first critical point is defined by
\a
4t_4 = -1, \qquad\qquad t_k =0 ~~~~k>4\0
\b
This implies in particular that $y = 4y_1$ and that the correlators will
be functions of $t_1,t_2$. Next, the degree analysis shows that
$a_1\sim t_2$ and $a_2 \sim t_1$. An elementary use of the first
eq.(\ref{Bouss}) shows that, up to an overall multiplicative
normalization constant, we can choose
\a
a_1 = 6t_2,\qquad \qquad a_2=3t_1 \label{a1a2'}
\b
This will be our choice (as it is consistent with the $W$ constraints
and the definition of the critical point). Now it is relatively easy to
use the recursion relations of the flow equations to compute the
correlation functions.

{\bf Proposition 5.3} {\it The exact one--point correlators
of $\widehat{\cal M}_3$ are
\a
<\tau_{3n-2+\epsilon}> &=& \sum_{h=0}^\infty <\tau_{3n-2+\epsilon}>_h \0\\
<\tau_{3n-2+\epsilon}>_h&=&
\sum_{\stackrel {j=0}{n-j+\epsilon\in 2{\bf Z}}}^{n - \frac{8h-2-\epsilon}{3}}
\frac{1}{48^h h!} \frac{2^{{{3n-3j+2+\epsilon}\over 2}-3h}
(-1)^{{{n-j+\epsilon}\over 2}-h}}
{3^{n-j-2+\epsilon} (n-1+\epsilon)!}
\frac{(3n-3j+2+\epsilon-6h)!!}{(3n-3j+2+\epsilon-8h)!!}\cdot\0\\
&& \cdot \frac{(3n-3+2\epsilon)!!!~(3n-2+2\epsilon)!!!}
{(3j)!!!~({{3n-3j+\epsilon}\over 2}-3h)!!!~
({{3n-3j+2+\epsilon}\over 2}-3h)!!!}
{}~t_1^j ~t_2^{{{3n-3j+2+\epsilon}\over 2}-4h}\label{taunhB}
\b
where $\epsilon=0,1$ and $n!!!$ is the 3 by 3 factorial,
i.e. $n!!! = n(n-3)(n-6)\ldots$ up to either $1 ~{ or}~ 2 ~{ or}~3$. By
convention $0!!! = (-1)!!! =1$, and ${1\over {n!!!}} = {1\over{n!!}} =0$
for $n\leq -2$. As a consequence in the above formula the exponents of
$t_1$ and $t_2$ are always non--negative.}

Sketch of proof. One has to remark first that, while the contributions from
two contiguous genera differ by $8y_1$, the recursion operators $D_{GG}$,...,
$D_{FF}$ contain contributions that differ by $4y_1$. It follows that
in $G_{r}$, $F_r$ there will appear {\it half--genus} contributions. Therefore,
at the first critical point, we have to start from the ansatz
(case $\epsilon =0$)
\a
G_{3n+1} &=& \sum_{s=0}^\infty G_{3n+1}^{(s/2)}, \qquad G_{3n+1}^{(s/2)} =
\sum_{\stackrel{j=0}{n-j \in 2{\bf Z}+1}} \al_j(n, s/2) ~t_1^j~
t_2^{{{3n-3j-1}\over 2}-2s}\label{ansG}\\
F_{3n+1} &=& \sum_{s=0}^\infty F_{3n+1}^{(s/2)}, \qquad F_{3n+1}^{(s/2)} =
\sum_{\stackrel{j=0}{n-j \in 2{\bf Z}}} \beta_j(n, s/2) ~t_1^j~
t_2^{{{3n-3j}\over 2}-2s}\label{ansF}
\b
where $s$ is the half--genus label, i.e. $s = 2h$, and the exponents of $t_1$
and $t_2$ are always non--negative. The half--genus contributions must not
appear in the correlators, thus we must have the physicality conditions
\a
\al_j(n,s/2) =0,\qquad 2\beta_j(n,s/2) = (j+1)\al_{j+1}(n, {{s-1}\over 2}),
\qquad s\in 2{\bf Z}+1 \label{physicality}
\b
Plugging (\ref{ansF}) and (\ref{ansG}) into (\ref{recurGr},\ref{recurFr}),
and using (\ref{physicality}) we find the following recursion relations for
the coefficients
\a
3j \al_j(n+1,h) = 3(3j-1) \al_{j-1}(n,h) + 12 j\beta_j(n,h)
+ 2j(j+1)(j+2) \beta_{j+2}(n,h-1) \label{recuralfa}
\b
for $ n-j\in 2{\bf Z}$, and
\a
3j \beta_j(n+1,h) &=& 3(3j-2) \beta_{j-1}(n,h) -24 j\al_j(n,h)
- 5j(j+1)(j+2) \al_{j+2}(n,h-1)\0\\
&& -{1\over 6} j(j+1)(j+2)(j+3)(j+4) \al_{j+4}(n, h-2) \label{recurbeta}
\b
for $ n-j\in 2{\bf Z}+1$.
These relations can be integrated and give the following result:
\a
\al_j(n,h)&=&\frac{1}{48^h h!} \frac{2^{{{3n-3j-1}\over 2}-3h}
(-1)^{{{n-j-1}\over 2}-h}}
{3^{n-j-1} (n-1)!} \frac{(3n-3j-1-6h)!!}{(3n-3j-1-8h)!!}\cdot\0\\
&& \cdot \frac{(3n-3)!!!~(3n-2)!!!}{(3j)!!!~({{3n-3j-1}\over 2}-3h)!!!~
({{3n-3j-3}\over 2}-3h)!!!},\qquad n-j\in 2{\bf Z}+1 \0
\b
and
\a
\beta_j(n,h)&=&\frac{1}{48^h h!} \frac{2^{{{3n-3j}\over 2} -3h}
(-1)^{{{n-j}\over 2}-h}}
{3^{n-j-1} (n-1)!} \frac{(3n-3j-6h)!!}{(3n-3j-8h)!!}\cdot\0\\
&& \cdot \frac{(3n-3)!!!~(3n-2)!!!}{(3j)!!!~({{3n-3j}\over 2} -3h)!!!~
({{3n-3j}\over 2}-3h-2)!!!} , \qquad n-j\in 2{\bf Z}\0
\b

Now we recall that
\a
<\tau_r\tau_1>_h = 3 G_{r+3}^{(h)},\qquad
<\tau_r\tau_2>_h = 6 F_{r+3}^{(h)}\0
\b
and integrate w.r.t. $t_1$ and $t_2$ the first and second expressions,
respectively. Comparing the results we find (\ref{taunhB}).
Just as in the KdV case we have to treat separately the case when both
exponents
of $t_1$ and $t_2$ in (\ref{taunhB}) vanish. The coefficients given by
(\ref{taunhB}) are the correct ones, but they have to be checked by means
of the W constraints.

Likewise we can compute $<\tau_{3n-1}>$ (case $\epsilon=1$).

The effective $W$ constraints in the case of the $\widehat{\cal M}_3$ model are
found once again by requiring that they be consistent with the Boussinesq flows
and that the $W$ generators form a closed algebra.

{\bf Proposition 5.4}. {\it The effective $W$ constraints
for the $\widehat{\cal M}_3$ model are
\a
L_n^{[r]}~Z^{1/3}=0, \qquad\quad r=1,2,\quad n\geq -r\label{Wbouss}
\b
where
\a
L^{[1]}_n &=& {1\over 3} \sum_{k=1}^\infty kt_k \frac {\d}{\d
t_{k+3n}}
 + {1\over {6}} \sum_{\stackrel {k,l}{k+l=3n}}
\frac {\d^2}{\d t_k \d t_l}
+{1\over 6}\sum_{\stackrel {k,l}{k+l=-3n}}kl t_kt_l +{1\over 9} \delta_{n,0}
\0,\quad
\forall n\0
\b
\a
{\sqrt 3 }L^{[2]}_n &=& {1\over 9}
\sum_{l_1,l_2=1}^\infty l_1l_2t_{l_1}t_{l_2}
\frac {\d}{\d t_{l_1+l_2+3n}}+
{1\over {9}} \sum_{\stackrel{l,k,j}{ l-k-j =-
3n}} lt_l\frac {\d^2}{\d t_k \d t_j}\0\\
&&+ {1\over {27}}
\sum_{\stackrel{l,k,j}{l+k+j=3n}}
\frac{\d^3}{\d t_l \d t_k \d t_j} +
{1\over 27}
\sum_{\stackrel{l,k,j}{l+k+j=-3n}}
lkj t_lt_kt_j,
\quad  \forall n\0
\b
In these expressions summations are limited to
the terms such that no index involved is either
negative or multiple of 3.
The above two sets of generators form a closed algebra,
the $W_3$ algebra,
\a
\relax [L^{[1]}_n, L^{[1]}_m] &=& (n-m)L^{[1]}_{n+m}
+\frac{1}{6} (n^3 -n) \delta_{n+m,0}\0\\
\relax [L^{[1]}_n, L^{[2]}_m] &=& (2n-m)L^{[2]}_{n+m}\0\\
\relax [L^{[2]}_n, L^{[2]}_m] &=& -\frac{1}{54}(n-m) \Big(
(n^2 + m^2 + 4nm) + 3 (n+m) +2\Big)L^{[1]}_{n+m} \0\\
&&+ \frac{1}{9} (n-m) \Lambda_{n+m} + {1\over {810}} n(n^2-1)(n^2-4)
\delta_{n+m,0}\0
\b
where
\a
\Lambda_n = \sum_{k\leq -1}L^{[1]}_kL^{[1]}_{n-k}
+ \sum_{k\geq 0}L^{[1]}_{n-k}L^{[1]}_k\0
\b
This corresponds to the quantum $W_3$ algebra with central charge 2.}

Sketch of proof. One can prove the consistency of the Boussinesq flows
with the above constraints in the following way. Call $K_n= Z^{-1/3}L^{[1]}_n
Z^{1/3}$. Then, for example, one can check that
\a
{1\over 2} D_{GF} \frac{\d K_n}{\d t_2} +D_{GG} \frac {\d K_n} {\d t_1} =
\frac {\d^2 K_{n+1}}{\d t_1^2} \0
\b
and so on. In fact it is not necessary to prove such equalities for any
$n\geq -1$, we simply need to do it for the first few cases, for
$K_n=0$ for $ n=-1,0,1,2$ implies $K_n=0, \forall n\geq-1$ due to the Virasoro
algebra structure. The same can be done for the higher order constraints.
Constants and polynomials in the couplings which appear in the generators
can be fixed by simply requiring algebraic closure.

\subsection{Other models}

Let us generalize what we have just done for the Boussinesq hierarchy
to the $\widehat{\cal M}_p$ (or p--th KdV) models.
The general recipe is as
follows. One must first of all disregard all the $t_k$ with $k$ a
multiple of $p$; the first critical point is
\a
(p+1) t_{p+1} = b, \quad\quad t_k=0 \quad\quad k>p+1\label{cpKdV}
\b
where $b$ is any number. The degree assignment is
\a
\relax [t_k]= p+1-k, \quad\quad [F^{(h)}]= (2 p+2)(1-h),\qquad
[a_{1}^{(0)}]=2, \ldots, [a_{p-1}^{(0)}] = p\label{degassKdV}
\b
where $a_{i}^{(0)}$ is the genus 0 part of $a_i$ and we have set, for
simplicity
$y_1=1$.
The CF's will be homogeneous functions of $t_1,\ldots, t_{p-1}$, which
constitute the small phase space.

In all the cases the method to compute CF's is the same as before.
We do not have however to redo literally the same steps as before. A shortcut
consists of fixing the form of the fields
by means of effective $W$--constraints, which in turn are determined imposing
compatibility with the relevant flow equations. Once this is done the CF's
can be obtained from the flow equations.

We write down hereafter the $W$--constraints for the general
$\widehat{\cal M}_p$
model. The $W$ constraints are
\a
L_n^{[r]} Z^{1/p} =0, \qquad \quad r=1,\ldots p-1,\qquad n\geq -r\label{Wcp}
\b
Compact formulas for the above generators can be written down by means of the
bosonic formalism. Let us introduce the current
\a
J(z) =\sum_{r=1}^\infty rt_r z^{r-1} + {1\over p} \sum_{r=1}^\infty z^{-r-1}
\frac{\d}{\d t_r}\0
\b
Then
\a
L^{[r]}_n = \frac{1}{p^{r-1}}{\rm Res}_{z=0} \Big( L^{[r]}(z) z^{pn+r}\Big),
\qquad L^{[r]}(z) = {1\over {r+1}}:J(z)^{r+1}:\0
\b
The normal ordering in the last definition is  the one between $t_r$ and
${\d \over {\d t_r}}$. The derivative is always supposed to stay at the right.
These generators close over a $W_p$ algebra with central charge $p-1$.
In particular we have
\a
L_n^{[1]} = \frac {1}{p} \sum_{k=1}^\infty kt_k \frac{\d}{\d t_{k+pn}}
+ {1\over{2p}} \sum_{k+l= pn}\frac{\d^2}{\d t_k \d t_l}
+ {1\over {2p}} \sum_{k+l=-np}klt_kt_l +\frac{p^2-1}{24 p} \delta_{n,0}\0
\b
In the above formulas
$n$ is any integer, and multiples of $p$ as well as non--positive integers
are excluded among the summation indices.

We can extract particular exact formulas as follows: we write down the
dispersionless
version of the constraints $L^{[r]}_{-r} Z^{1/p}=0$, with $r=1,\ldots,p-1$;
this equation gives a recursion relation for $<\tau_r>$, $r= 1,\ldots,
p-1$, in terms of
$<\tau_l>$ with $l<r$, which can be solved and gives:
\a
<\tau_r> &=& \sum_{n=1}^{p-1}(-1)^{n} \prod_{i=1}^{n-1}
 \sum_{s_i=1}^{\frac{r_i-1}{2}} b^{-s_i} \left(\bac r_i \\ s_i\ea\right)
 \sum_{\stackrel{l_1^{(i)}, \ldots,l_{s_i}^{(i)}}
{l_1^{(i)}+ \ldots+l_{s_i}^{(i)} +r_i -s_i(p+1) = r_{i+1}}}
l_1^{(i)} \ldots l_{s_i}^{(i)}t_{l_1^{(i)}}\ldots t_{l_{s_i}^{(i)}}\cdot\0\\
&& \cdot\sum_{s_n=0}^{r-2} b^{-r+s_n} \left(\bac r+1 \\ s_n\ea\right)
 \sum_{\stackrel{l_1^{(n)}, \ldots,l_{r-s_n+1}^{(n)}}
{l_1^{(n)}+ \ldots+l_{r-s_n+1}^{(n)} +s_n(p+1) = pr}}
l_1^{(n)} \ldots l_{r-s_n+1}^{(n)}t_{l_1^{(n)}}
\ldots t_{l_{r-s_n+1}^{(n)}}\label{taurp}
\b
Although this result has been obtained from the a genus 0 approximation, {\it
it
is an exact result}. In particular, setting $b=-1$, we have
\a
<\tau_1> = {p\over 2} \sum_{k,l; k+l=p} klt_kt_l \label{tau1p}
\b
As a consequence
\a
<\tau_1\tau_k\tau_l> = pkl\delta_{l,p-k},\qquad 1\leq k,l\leq
p-1\label{metricp}
\b
which specifies the metric of the corresponding topological field theory,
\cite{BX7}.

\subsection{Higher critical points}

In all the previous examples the first critical point
has been characterized by a dependence of the basic
fields on the couplings specified by homogeneous polynomials with
non--negative integer powers. Higher critical points are characterized
still by a homogeneous dependence, but with rational and/or negative
powers of the couplings.

The procedure to compute correlation functions is always the same: a shortcut
to arrive at the results is to use simultaneously W constraints and flow
equations. With a good deal of perseverance we could probably arrive at
exact correlators as in the previous subsections. However,
in order to give an idea and for future reference,
we think it is enough to present a few partial results.

Let us start with the KdV model. The (Kazakov's) critical points were defined
in subsection 5.1. At these points the degree assignments (setting $y_1=1$) are
\a
\relax [t_l]= 2k+1-l, \quad\quad [F^{(h)}]= 2 (2k+1)(1-h) ,\quad\quad
[a^{(h)}]=2-(2k+1)2h\label{hdegassKdV}
\b
Indeed, contrary to the first critical point, we have to expect
non--vanishing contributions from all genera for the field $a= a^{(0)}+ a^{(1)}
+a^{(2)}+...$. We find the following results
\a
a^{(0)}&=& \al(k)^{1/k} t_1^{1/k} ,\qquad\qquad \al(k) =
\frac{k!}{2^{k-1}(2k-1)!!}
\0\\
a^{(1)}&=& \beta(k) t_1^{-2}, \qquad\qquad \beta(k)= {1\over{24}} \frac
{k-1}{k}
\0\\
a^{(2)}&=& -\al(k)^{-{1/k}}\gamma(k)t_1^{-\frac{4k+1}{k}}\0
\b
where
\a
\gamma(k) = \frac{1}{720k^3} (k-1)(32k^3-72k^2 +177 k - 77)\0
\b
Knowing this formulas one can calculate the correlators in genus 0,1
and 2. The expression for $a^{(1)}$ is also in ref.\cite{DS}

Higher critical points of more complicated models can be found in
\cite{BX7}

\centerline{-------------------------}

The models $\widehat {\cal M}_p$ have a topological field theory
interpretation.
The corresponding topological field theories are easily identified with
those of the A series in the ADE classification \cite{BX7}. The latter are
known to be based on p--KdV hierarchies. Therefore, what we have achieved
in this section is a new presentation of this old subject, in fact a very
powerful presentation since it has allowed us to calculate new all--genus
expressions for the correlators. However, if we look at this section not
from the point of view of topological field theories but from the matrix
model point of view we cannot yet be satisfied. Although we have used
the reduced integrable hierarchies obtained in section 4 from two--matrix
models, and we have used other matrix model inputs, a direct connection between
two--matrix models and the results found in this section, although
very plausible, has not yet been established: in particular some
normalization constants have been
arbitrarily fixed and the identification $a_1\equiv <\tau_1\tau_1>$ deserves
a safer ground.

The purpose of the next section is to provide such connection.

\section{Constrained two--matrix models.}

In this section we want to introduce and analyze constrained two--matrix
models that are characterized by p--KdV hierarchies.
We are going to study in particular detail the ones based on KdV and
Boussinesq hierarchy. In the previous two sections we learned that, in order to
end up with the p--th KdV hierarchy, we have to impose the constraints
$S_i=0$, which amounts in the Toda lattice formalism to impose that
the diagonal elements of the matrix $Q(1)$ vanish identically.
We will just impose this constraint on the model ${\cal M}_{p_1,p_2}$, and call
the two--matrix models so obtained ${\cal M}_{p_1,p_2}^{(0)}$. The main
result of this section is that the reduced models
$\widehat {\cal M}_p$ can be imbedded
in the {\it renormalized} ${\cal M}_{p+1,p}^{(0)}$.

A remarkable aspect of our derivation is that {\it we obtain all the results
via the flows of the Toda lattice hierarchy and we never abandon the framework
of two--matrix models}. A byproduct of this section is a new way to derive
p--KdV hierarchies from Toda lattice hierarchies.

\subsection{The constrained two--matrix model ${\cal M}_{3,2}^{(0)}$ and the
KdV hierarchy}.

The simplest interesting model is ${\cal M}_{3,2}^{(0)}$. Let us start
from the coupling conditions of the model ${\cal M}_{3,2}$ of section 3.2,
and impose the condition $a_0(n)=0$.

\subsubsection{The coupling conditions in ${\cal M}_{3,2}^{(0)}$}

Remember that KdV, at the first critical point, is expected at
$t_2=0$ and $6t_3= -1$. Imposing this and
\a
a_0(n) =0\label{important}
\b
eqs.(\ref{coum32}) becomes, in genus 0,
\a
&&2cb_2 -R^2=0, \qquad  b_1=0,\qquad a_1=t_1-\frac{cs_1}{2s_2}\0\\
&&R=-{n \over c},\qquad b_0(n) = - \frac{s_1}{2 s_2},\qquad
2s_2R+ca_1=0\label{coum320}
\b
Setting for simplicity $s_1=0$, one gets in particular
\a
a_1 =t_1 \label{initcond}
\b

The importance of (\ref{initcond}) is that, from the constrained coupling
conditions, we have obtained $a_1 =t_1$, a result which was postulated in
section 5.1 on the basis of plausibility arguments.

{\it Remark 1}. The particular values chosen for $s_1, t_2,t_3$ are not
important. Other choices would not change qualitatively
the results, but only renormalize them either additively ($s_1$)
or multiplicatively ($t_3,t_2$).

{\it Remark 2}. The last equation (\ref{coum320}) implies
\a
t_1 = \frac{2s_2}{c^2} n \label{dyncond}
\b
The conditions (\ref{initcond}) and (\ref{dyncond}) have
to be imposed as a last condition on the correlators after all the
calculations have been carried out (see below),
therefore the fact that $t_1\sim n$ does
not interfere with differentiating or integrating with respect to $n$. The
rather mysterious condition (\ref{dyncond})
may have an interesting topological field theory interpretation, see
\cite{BX8}.

{\it Remark 3}. We stress that it is irrelevant whether $a_1=t_1$ is the unique
solution of the coupling conditions and that it be found in genus 0. What is
important is that we be able to impose it (at every genus) without breaking
integrability. This is in fact what we are going to show next.

\subsection{KdV flows from Toda flows}.

We show next that we can extract the KdV flows directly from the Toda
lattice hierarchy. We recall that in ${\cal M}_{3,2}$
\a
Q(1) = I_+ + \sum_i\Big(a_0(i)E_{i,i}+ a_1(i)E_{i,i-1}\Big)\0
\b
Setting $Q\equiv Q(1)$, we have

{\bf Lemma 6.1}. {\it In ${\cal M}_{3,2}$ the formulas, obtained from
the Toda lattice hierarchy and from eq.(\ref{ddZ})
\a
<\tau_1\tau_k> = \tr \Big([Q_+,Q^k]\Big)\label{tau1tauk}
\b
give rise to the flows of the NLS hierarchy.}

Proof. We notice that, due to eq.(\ref{Fa1}), we have
\a
\frac{\d a_1}{\d t_k} = \frac {\d}{\d t_1}<\tau_1\tau_k>\0
\b
This is the first ingredient. The second ingredient are the first flows
(\ref{f11}), which, in the ${\cal M}_{3,2}$ case, take the form
\a
S_1' = (1-D_0^{-1})a_1,\qquad\quad a_1'= a_1(D_0-1)S_1\label{ffnls}
\b
(remember that we defined $S_1(n)= a_0(n-1)$).
Here we have introduced a new notation, which turns out to be very convenient
in this kind of trade. For any discrete function $f_N$ we define
\a
(D_0 f)_N = f_{N+1}\0
\b
We will also use the notation $e^{\d_0}$ instead of $D_0$,
with the following difference: $D_0$ applies to the nearest right neighbour,
while $e^{\d_0}$ is meant to act on whatever is on its right.

Using this notation we can write $Q$ as follows
\a
Q = e^{\d_0} + D_0S_1 + a_1 e^{-\d_0}\label{Qnls}
\b
We also remark that the sum, $\sum_{n=0}^{N-1}$, in $\tr$ is exactly the
inverse of the operation $D_0-1$.
Now it remains for us to evaluate the RHS of (\ref{tau1tauk}).
Hereon we give an example; a more complete proof will be provided
elsewhere
\a
\frac {\d a_1}{\d t_3}&=& \frac {\d}{\d t_1}\tr \Big([Q_+, Q^3]\Big)\0\\
&=&\frac{\d}{\d t_1}\Big(D_0a_1a_1 +a_1 D_0^{-1}a_1 +a_1a_1 +(D_0S_1)^2a_1
+S_1^2a_1 +a_1S_1D_0S_1\Big)\0\\
&=&\Big(a_1'' + 3a_1^2+ 3S_1^2a_1 + 3S_1a_1'\Big)'\0
\b
which is exactly the third NLS flow for $a_1$. In order to find the
flows for $S_1$ one has simply to differentiate
\a
S_1 = \d^{-1}(1-D_0^{-1})a_1\0
\b
and use again the first flows.

Using this Lemma, it is now easy to conclude our argument. We pass to
the ${\cal M}_{3,2}^{(0)}$ model setting
$S_1=0$. The odd flows of $a_1$ reduce to the KdV flows in exactly the same
form given by the recursion relations (\ref{recurKdV}). Not only are
these relations compatible with the W constraints of section 5.2, but also
the result $a_1=t_1$, obtained from the coupling conditions, is, and coincides
with our assumption (\ref{ansa}). {\it We conclude that the model
$\widehat{\cal M}_2$ is imbedded in the constrained two--matrix model
${\cal M}_{3,2}^{(0)}$, and that all the results obtained in section 5.1,5.2
hold true for the latter}.

{\it Remark.} The second sector of ${\cal M}_{3,2}^{(0)}$, i.e. the dependence
on $s_k$, as well as the dependence on the bilinear coupling $c$, does not play
a role in the above arguments. They can at most renormalize the final results
(as noticed above). In other words, the second sector is a spectator. Whether
and how it is possible to compute correlators of the second sector, i.e.
correlators of $\sigma_k$, or mixed ones, is a question that we leave open
here.

\subsection{The k--th KdV critical point and ${\cal M}_{2k+1,2}^{(0)}$}

To confirm the result just obtained let us look at the k--th KdV critical
point (section 5.5). This critical point turns out to be embedded in the
model ${\cal M}_{2k+1,2}^{(0)}$. The relevant potentials in this case are
\a
V_1(\lambda_1) = t_{2k+1} \lambda_1^{2k+1} +t_1\lambda_1,
\qquad\qquad V_2(\lambda_2) = s_2 \lambda_2^2 +s_1\lambda_2\0
\b
which entail the following coupling conditions
\a
P^\circ (1) +(2k+1)t_{2k+1} Q(1)^{2k} +t_1 + c Q(2) =0 ,\qquad
\bar{\cal P}^\circ (2) +2s_2 Q(2) +s_1 +cQ(1) =0\0
\b
The relevant equations one gets in genus 0 are
\a
&& (2k+1)t_{2k+1} \left(\bac 2k\\ k \ea\right) a_1^k + c b_0 +t_1 =0\0\\
&&2s_2 b_0 +s_1 =0, \qquad 2s_2 R + c a_1 =0,\qquad2s_2 b_1 + c R +n =0 \0
\b
One finds
\a
b_0 = - \frac{s_1}{2s_2},\qquad\qquad a_1 =
\Big(- \frac{1}{(2k+1)t_{2k+1}} \left(\bac 2k\\k\ea\right)^{-1}
(t_1- \frac{s_1}{2s_2}) \Big)^{1/k} \0
\b
Here, either we make the replacement $t_1 \rightarrow \tilde
t_1= t_1- \frac{s_1}{2 s_2}$
or simply set $s_1=0$. Moreover we set (k-th critical point)
$2(2k+1)t_{2k+1}=-1$. Then
\a
a_1 = \Big( \frac{k!}{2^{k-1}(2k-1)!!} t_1\Big)^{1/k}\label{kthcriti}
\b
This result coincides exactly with the analogous formula in section 5.5.
Substituting this in the flows we can calculate the KdV
correlators for any critical point in genus 0.

\subsection{The renormalized two--matrix  model ${\cal M}_{4,3}^{(0)}$ and the
Boussinesq hierarchy}

We want now to repeat the same for the Boussinesq hierarchy. We have to start
from the ${\cal M}_{4,3}^{(0)}$, but one soon realizes that things are not
as simple as in the previous example. In fact an analog of Lemma 6.1 holds,
but setting $a_0=0$ in the resulting flows does not lead to the Boussinesq
flows. A significant change of strategy is necessary. But let us start once
again from the coupling conditions for ${\cal M}_{4,3}$.

\subsubsection{The coupling conditions in ${\cal M}_{4,3}^{(0)}$}

The coupling conditions for the two--matrix model ${\cal M}_{4,3}$ are
\a
&& P^\circ(1)+4t_4 Q(1)^3 + 3t_3 Q(1)^2 + 2t_2 Q(1)  + t_1 + c Q(1) =0\0\\
&& \bar {\cal P}^\circ_2 +3s_3 Q(2)^2+ 2s_2 Q(2) + s_1 + c Q(1)
=0\label{coum43}
\b
These can be expressed in terms of the fields $a_0, a_1, a_2, b_0,
b_1, b_2, b_3$ and $R$.
We expect to find the Boussinesq hierarchy for $a_0=0$ and $t_3=0$.
The coupling conditions in genus 0 then become
\a\baccc
cb_3 = - 4t_4 R^3,& cb_2 =0,&cb_1=- 2t_2 R - 12t_4a_1 R,\\
cb_0 = - t_1  - 12t_4a_2,&n + 12 t_4 a_1^2 + 2t_2 a_1 + cR =0,
&3s_3 R^2 + c a_2 =0,\\
6s_3 b_0R + 2s_2R + c a_1=0,&s_3(2b_1 +b_0^2) + 2s_2 b_0 +s_1=0,
&6s_3b_0b_1 + 2s_2 b_1 + c R +n = 0\ea \0
\b
For simplicity we choose $b_1=0$ and $s_1=0$, which leads to
\a
a_1 = - \frac{t_2}{6t_4},\qquad a_2= -\frac{t_1}{12t_4},
\qquad R= -\frac{n}{c}\label{initm43}
\b
After determining $b_0=0$ and $b_3$ we are left with two conditions of the type
(\ref{dyncond}) among the couplings, which are irrelevant for the following
developments. Different choices of $b_1$ and $s_1$ would simply imply additive
redefinitions (renormalizations) of $t_1$ and $t_2$, therefore we ignore them.

{\it It is important that up to a global rescaling (see below) we have found
the same results that were assumed under (very) plausible arguments in section
5.3, see eq.(\ref{a1a2'}).}

\subsubsection{Boussinesq flows from Toda flows}

As we have anticipated above, in the ${\cal M}_{4,3}$ there is an analog
of Lemma 6.1. In particular, from the Toda flows we can obtain the flows of
the four--field KP hierarchy. However this is irrelevant to our problem,
since, setting $S_1=S_2=0$ does not lead to the Boussinesq hierarchy (or to
any integrable hierarchy, for that matter). The reason is well--known: the
above conditions oblige the system to flow outside the manifolds of
the flow equations. To preserve integrability we have to introduce in the
original hierarchy a (presumably infinite) set of corrections. In field theory
language we can say that the constraint $a_0=0$ can be imposed without
spoiling integrability only at the price of introducing a (presumably infinite)
set of counterterms. The very important point is that this set of counterterms
can be exactly computed, after which the resulting model, referred to
henceforth as the ${\cal M}_{4,3}^{(0,r)}$ model, will accomodate the
Boussinesq hierarchy.

The recipe to obtain the result is as follows:

1) Define the general matrix
\a
\widehat Q = e^{\d_0} + \sum_{i=1}^{\infty}\hat a_i e^{-i\d_0}\label{hatQ}
\b

2) Assume as first flows
\a
D_0\hat a_1=\hat a_1, \qquad\qquad D_0\hat
a_i = \hat a_i + \hat a_{i-1}', \quad i\geq 2\label{ffB}
\b
(these are the usual first flows in which we have set $\hat a_0=0$).

3) Now impose
\a
\widehat Q^3 = e^{3\d_0} + 3a_1e^{\d_0} + 3a_2\label{Bcond}
\b
This equation and (\ref{ffB}) completely determine $\hat a_i$ in terms
of $a_1,a_2$,
\a
&&\hat a_1 =\bar a_1\equiv a_1,\qquad \hat a_2 = \bar a_2\equiv a_2 - a_1',
\qquad \hat a_3 =\bar a_3\equiv -a_2'-a_1^2+\frac{2}{3} a_1'',\0\\
&&\hat a_4 =\bar a_4\equiv {2\over 3} a_2'' - {1\over 3} a_1''' +
4 a_1a_1' - 2 a_1 a_2\0
\b
and so on. $\widehat Q$ is now to replace the matrix $Q(1)$ of the model
${\cal M}_{4,3}$ with $a_0=0$. It contains the right counterterms to
generate the Boussinesq hierarchy. To this end,

4) use the following

{\bf Lemma 6.2} {\it Replace $Q(1)$ with $\widehat Q$ in the Toda lattice
hierarchy formulas and evaluate them at $\hat a_i=\bar a_i$
\a
\frac {\d a_1} {\d t_k} = \frac {\d}{\d t_1}
\tr \Big([\widehat Q_+, \widehat Q^k]\Big)|_{\hat a_i=\bar a_i},
\qquad k\neq 3n\label{Bkfs}
\b
and the like. Then these formulae provide a realization of the
Boussinesq hierarchy.}

Proof. We limit ourselves to a few examples. A more complete proof will
be given elsewhere.
\a
&&\frac {\d a_1} {\d t_2} = \frac {\d}{\d t_1}
\tr \Big([\widehat Q_+, \widehat Q^2]\Big)|_{\hat a_i=\bar a_i}=
(D_0\tilde a_2 + \tilde a_2)'|_{\hat a_i=\bar a_i} =
2a_2' -a_1''\0\\
&&\frac {\d a_1} {\d t_4} = \frac {\d}{\d t_1}
\tr \Big([\widehat Q_+, \widehat Q^4]\Big)|_{\hat a_i=\bar a_i}=
{2\over 3} a_2'' - {1\over 3} a_1'''
-2 a_1a_1' +4a_1a_2 \0\\
&&\frac {\d a_2} {\d t_2} = {1\over 2}\frac {\d}{\d t_1}
\tr \Big([\widehat Q_+^2, \widehat Q^2]\Big)|_{\hat a_i=\bar a_i} +
{1\over 2} \frac {\d a_1}{\d t_2}
= a_2'-a_1^2 -{2\over 3}a_1''\0
\b
and so on. These are the Boussinesq flows, even though not in exactly the same
form as in section 4.1 and 5.3. In fact we have to multiply by 3 the $a_1,a_2$
fields there to obtain the flows here.

{\it Remark 1}. We should have allowed also for a field $a_0$ in $\tilde Q$
and set it to zero at the end of the calculations. This could of course
have been done
but would not have changed the results. Setting $a_0=0$ from the very
beginning we have
simply anticipated the result and simplified the formulas a lot.

{\it Remark 2}. It is not surprising that we have found some disagreements
in the normalizations here compared sections 4 and 5. We have already pointed
out at the end of section 5 that some identifications made there were
likely to be arbitrary as far as the normalizations are concerned.

Up to the normalization problem illustrated in the previous remark we can see
that {\it the flow equations, the coupling conditions (\ref{initm43}) and
consequently the W constraints pertinent to the $\widehat {\cal M}_3$ model
can be embedded in the ${\cal M}_{4,3}^{(0,r)}$ matrix model}. If we set the
critical point at $12t_4=-1$ in (\ref{initm43}) and multiply by 3 the fields
$a_1,a_2$ in section 4 and 5, we can simply transfer the results obtained
there to ${\cal M}_{4,3}^{(0,r)}$. To be more precise, we summarize the
results concerning the latter as follows:

{\it i)} The Boussinesq flows are given by
\a
\frac{\d a_1}{\d t_r} = G_{r+3}'
&=& D_{GG} G_r + D_{GF}F_r\label{recurGr'}\\
\frac{\d a_2}{\d t_r} = F_{r+3}'
&=& D_{FG}G_r + D_{FF}F_r\label{recurFr'}
\b
with $F_1=1, ~G_1 =0$ and $F_2=0,~G_2=1$.
The differential operators are
\ai
&&D_{GG} = 3a_2 \d +2a_2' - a_1 \d^2 - 2 a_1' \d - a_1''
-{1\over 3}\d^4\label{DGG'}\\
&&D_{GF} = {2\over 3}\d^3 + 2 a_1 \d + a_1'\label{DGF'}\\
&&D_{FG} = 2a_2'\d + a_2'' - 2\Big(a_1 a_1' + a_1^2\d + {2\over 3} a_1\d^3
+ {1\over 3}a_1''' + a_1''\d +  a_1'\d^2 +{1\over 9} \d^5\Big)\label{DFG'}\\
&&D_{FF} ={1\over 3} \d^4 + a_1 \d^2 + a_2' + 3a_2\d \label{DFF'}
\bj

{\it ii)} The coupling conditions imply, at $12t_4=-1$,
\a
a_1 = 2t_2,\qquad\qquad a_2 =t_1\label{a1a2''}
\b

{\it iii)} The correlators are the same as in section 5.5 except for a global
factor of 3. In particular the RHS of (\ref{taunhB}) must be divided by 3.

{\it iv)} The $W$ constraints appropriate for ${\cal M}_{4,3}^{(0,r)}$
are
\a
L_n^{[r]}~Z=0, \qquad\quad r=1,2,\quad n\geq -r\label{Wbouss'}
\b
the generators being the same as in Proposition 5.4. But in order to
reproduce the right correlators we have to compute them at $4t_4=-1$. In other
words the location of the critical point gets renormalized.

\section{Conclusion}

The procedure introduced above for the Boussinesq hierarchy holds for
any p--KdV hierarchy. We can always define suitable coordinates
in the matrix $\widehat Q$,
which, inserted in the Toda lattice flows, generate the p--KdV flows. The
recipe is just a generalization of the one given above. This is certainly
a remarkable (and new, to our knowledge) result, which deserves further
elaboration.

Finally we can draw the following conclusion: the p--KdV hierarchies are
contained in specific {\it constrained} two--matrix models; once we impose
the constraint, the $Q$ matrix of the relevant model has to be suitably
redefined (except in the 2--KdV case) to insure integrability; the counterterms
can be exactly calculated and give rise to ``renormalized'' coordinates;
in turn these coordinates, when substituted in the formulas
of the Toda lattice hierarchy, gives rise to the p--KdV flows. The same
procedure may well be applicable to extract from two--matrix models  other
hierarchies such as those studied in \cite{AK}.

\subsection*{Appendix}

Here are the exact 2 point CF's of the model ${\cal M}_{2,2}$. Let us first
define the functions
\a
&& {\cal F}_{n,m}(\al_2, \beta_1,\beta_2, \gamma_1, \gamma_2)
=\sum_{l=0}^n \sum_{k=0}^{l/2} \sum_{r=0}^{\frac{l-2k}{2}} \sum_{p=0}^m
\sum_{q=0}^{p/2} \frac {n!m!}{2^{k+q} (n-l)!(m-p)! k! q!}\cdot\0\\
&&\cdot \sum_{j=0}^{min(l-2k,p-2q)} \Big( \frac{1}{r!j! (r +k-q+\frac{p-l}{2})!
(l-2k-r-j)! (\frac{l+p}{2} -k-q-r-j)!} \0\\
&&~~~~~~~~~~~~~~~~~~~~ - \frac{1}{j! (l-2k-r)! (r-j)! (\frac{l+p}{2} -k-q-r)!
(\frac{p-l}{2} -q+k+r-j)!} \Big) \cdot\0\\
&&\cdot \Big(\frac{l+p}{2} -k-q-j\Big)! \left( \bac N\\ \frac{l+p}{2} -k-q-j+1
\ea \right)~\al_2^{{{p-l}\over 2}+k+r}\beta_1^{n-l}\beta_2^{m-p} \gamma_1^{k+r}
\gamma_2^{{{p+l}\over 2}-k-r}\0
\b
where $\al_2, \beta_1, \beta_2,\gamma_1,\gamma_2$ were defined in
eq.(\ref{symbols}). Then
\a
<\tau_n \sigma_m> &=& \tr ([Q(1)^n, Q(2)^m_-]) =
{\cal F}_{n,m}(\al_2, \beta_1, \beta_2,\gamma_1,\gamma_2)\label{taunsigman}\\
<\tau_n \tau_m> &=& \tr ([Q(1)^n, Q(1)^m_-]) =
{\cal F}_{n,m}(1, \beta_1, \beta_1,\gamma_1,\gamma_1)\label{tauntaun}
\b
Finally $<\sigma_n \sigma_m> $ is obtained from $<\tau_n \tau_m>$ with
the exchange $t_k \leftrightarrow s_k$ for $k=1,2$.

\vskip.3cm
{\bf Acknowledgements} One of us (C.P.C.) would like to thank R.Paunov for
discussions, SISSA for hospitality and CNPq (Brasil) for financial support.

\end{document}